\documentclass[12pt,a4paper,dvips,final]{article}
\usepackage{rotating}
\usepackage{a4p}
\usepackage{cite}
\usepackage{graphicx}
\usepackage{physics }
\usepackage{l3_titlePH,ifthen}
\usepackage{amsfonts}
\usepackage{amsmath}
\def\d{{\rm d}}
\def\numev{74859} 

\def\numevz{7535}

\def\PL{{\it Phys. Lett. }}
\def\ZP{{\it Z. Phys. }}
\def\NIM{{\it Nucl. Inst. Meth. }}

\def\PR{{\it Phys. Rev. }}
\def\PRL{{\it Phys. Rev. Lett. }}
\def\ep{\mbox{e}^+}
\def\em{\mbox{e}^-}
 
\def\gaga{\gamma\gamma}

\newcommand{\piz}{\pi^0}
\newcommand{\rp}{\rho^+}
\newcommand{\rmi}{\rho^-}
\newcommand{\rz}{\rho^0}
\newcommand{\rprm}{\rp\rmi}
\newcommand{\rzrz}{\rz\rz}

\newcommand{\pppp}{\pip\pim\pip\pim}
\def\pppz{\pip\piz\pim\piz}
\def\roro{\rho\rho}
\newcommand{\ppzppz}{\pip\piz\pim\piz}
\newcommand{\eepppp}{\ep\em\ra\ep\em\pip\pim\pip\pim}
\newcommand{\eeppzppz}{\ep\em\ra\ep\em\pip\piz\pim\piz}
\newcommand{\ggpppp}{\gaga\ra\pppp}
\newcommand{\ggppzppz}{\gaga\ra\ppzppz}

\newcommand{\ggrprm}{\gaga\ra\rp\rmi}
\newcommand{\ggrzrz}{\gaga\ra\rz\rz}

\newcommand{\wgg}{W_{\gaga}}

\newcommand{\beq}{\begin{equation}}
\newcommand{\eeq}{\end{equation}}
\newcommand{\btab}{\begin{table}}
\newcommand{\etab}{\end{table}}
 
\preprint{2005-058}
\date{December 12, 2005}
\journalname{Physics Letters B}
\newlength{\capindent}
\newlength{\capwidth}
\newlength{\figwidth}
\newcommand{\icaption}[2][!*!,!]{\hspace*{\capindent}%
  \begin{minipage}{\capwidth}
    \ifthenelse{\equal{#1}{!*!,!}}%
      {\caption{#2}}%
      {\caption[#1]{#2}}
  \end{minipage}}
\begin{document}

\begin{titlepage}
\title{Analysis of the $\boldsymbol\pppp$ and $\boldsymbol\ppzppz$ 
Final States \\
in Quasi-Real Two-Photon Collisions at LEP}
\author{The L3 Collaboration}
%
%
\begin{abstract}
The reactions $\ggpppp$ and $\ggppzppz$ are studied with the L3
detector at LEP in a data sample  
collected at  centre-of-mass energies from $161\GeV$ to $209\GeV$ 
with a total integrated luminosity of 698~pb$^{-1}$.
A spin-parity-helicity
analysis of the  $\rz\rz$ and $\rp\rmi$  systems for  
two-photon centre-of-mass energies between $1\GeV$ and $3\GeV$ shows the dominance of 
the spin-parity state
$2^+$ with helicity 2. 
The contribution of $0^+$ and $0^-$ spin-parity states is also observed, whereas
 contributions of $2^-$ states and of a state with spin-parity
$2^+$ and zero helicity are found to be negligible.
\end{abstract}

\submitted

\end{titlepage}

\section{Introduction}
Several experiments have observed a large cross section near threshold for the reaction
$\ggrzrz$~\cite{TASSO,TASSO1,ARGUS}. 
In contrast,  the corresponding cross section for the isospin-related 
reaction $\ggrprm$ was shown to be small~\cite{ARGUS1, ARGUS0}.
The first spin-parity-helicity analysis of the reaction $\ggpppp$ 
was carried out by the TASSO Collaboration~\cite{TASSO1} by studying angular 
correlations.  The data sample consisted of 1722 
events for two-photon centre-of-mass energies 
 $1.2\GeV< \wgg  < 2.0\GeV $. A spin-parity-helicity
analysis with higher statistics was performed by the ARGUS Collaboration 
\cite{ARGUS} with 5181 events  in the  region 
$1.1\GeV<\wgg <2.3\GeV $.
Both collaborations used similar models and observed the dominance 
of  $\rz\rz$ states with spin-parity $J^P=2^+$ and $0^+$. 
The contribution of   negative-parity states was found to be 
negligible.

A number of theoretical models~\cite{Rosner} were proposed to interpret these experimental 
results. In a $t$-channel factorization  approach~\cite{Alex}, the
$\ggrzrz$ cross  section is related to photo-production and 
hadronic cross sections at low  energies. This model leads to the
interpretation of the broad enhancement in the $\ggrzrz$ cross section
around $1.6\GeV{}$ as a threshold behaviour due to Regge
exchange. Other models suggest an $s$-channel $\rz\rz$ resonance  
\cite{Li,Acha}, either a normal $\rm q\bar{q}$ state or a four-quark 
$\rm qq\bar{q}\bar{q}$ bound state. In four-quark models, isoscalar
and isotensor  
resonances interfere destructively to suppress the $\ggrprm$ signal and 
constructively to describe the  $\ggrzrz$ cross section. The
proposed models  differ substantially in the predicted cross section
for the  production of other vector mesons such as $\gaga \ra \rz\omega$ and
$\gaga\ra\phi\phi$. 

This Letter presents  the results of a spin-parity-helicity analysis of the
reactions $\ggpppp$ and $\ggppzppz$  in data collected by the L3 detector~\cite{L3D} at LEP, 
 using  the same technique as TASSO and ARGUS.
The data samples consist of $7.5 \times 10^4$ events for the $\eepppp$ 
channel and $7.5 \times 10^3$ events for the  $\eeppzppz$ channel.
These data are selected in the region of quasi-real photons  with a
 maximum virtuality of   $Q^2 \simeq 0.02 \GeV ^2$.
The   $\ggrzrz$ and $\ggrprm $ cross sections obtained in this analysis are 
compared to the high-virtuality~\cite{high00,highch}
and mid-virtuality~\cite{vsat00,vsatch} data obtained 
with the same detector.

\section{Data and Monte Carlo Samples}

The two-photon production of a $\rho$-pair, $\ggrzrz$ or $\ggrprm$, is
observed via the reactions $\eepppp$ or $\eeppzppz$, respectively. Detection of
the scattered leptons is not required.
The data were collected  with  the L3 detector at $\ee$
centre-of-mass energies $\sqrt s=161-209\GeV{}$, with a total
integrated luminosity  ${\mathcal{L}}_{\epem}=697.7$~pb$^{-1}$ and an average
centre-of-mass energy of 196$\GeV$.
 The analysis 
described in this paper is mainly  based on the central tracking
system and the electromagnetic calorimeter. 

Four-pion Monte Carlo  events are generated with the EGPC~\cite{EGPC}
program.
The four-momentum of the two-photon system is distributed according to the
transverse two-photon
luminosity function~\cite{Bud}. The pion
four-momentum vectors  are generated using four-particle phase
space. 
The  events are then passed through the L3 detector
simulation, which uses  
the GEANT~\cite{GEANT} and GEISHA~\cite{GEISHA} programs, and  
are reconstructed  
following the same procedure as used for the data. 

\section{Event Selection}

The events are collected by two charged-track triggers. The first
trigger~\cite{L3T} requires at 
least two wide-angle tracks, back-to-back within $\pm 41^\circ$ in the plane transverse to the
beam. The second trigger \cite{L3IT} is based on a
neural network which was
trained to select low-multiplicity events while rejecting beam-gas and
beam-wall background.

Events are selected by requiring:
\begin{itemize}
\item four charged tracks for the $\eepppp$ reaction and
two charged tracks for the $\eeppzppz$ reaction,
 with a net charge of zero in each case. A track is
required to have: more than
12 hits,  with at least 60\% of possible hits, 
a transverse
momentum, $p_\mathrm{t}$,  greater than 100$\MeV$ and a distance of closest
approach to the interaction vertex in the transverse plane less
than 2 mm.

\item no photons for the $\ggpppp$ reaction and four isolated
clusters in the electromagnetic calorimeter
for the $\ggppzppz$ reaction. A photon is defined as an
isolated shower  in the electromagnetic calorimeter consisting of
 at least two adjacent crystals with an energy
greater than $100\MeV{}$ and with no charged track within  200~mrad.

\item an energy loss  $\d E/\d x$ in the tracking chamber corresponding to the hypothesis   
that all the charged particles are  pions, with a confidence level
greater than 6\%. 

\item two pairs of photons each with a good fit to the  
$\pi^0$ decay hypothesis  for the $\pppz$ final state. 
\end{itemize}

\begin{figure}[htbp]
\begin{center}
\halign{&\includegraphics[height=0.37\textheight,width=.5\textwidth]{#}\cr
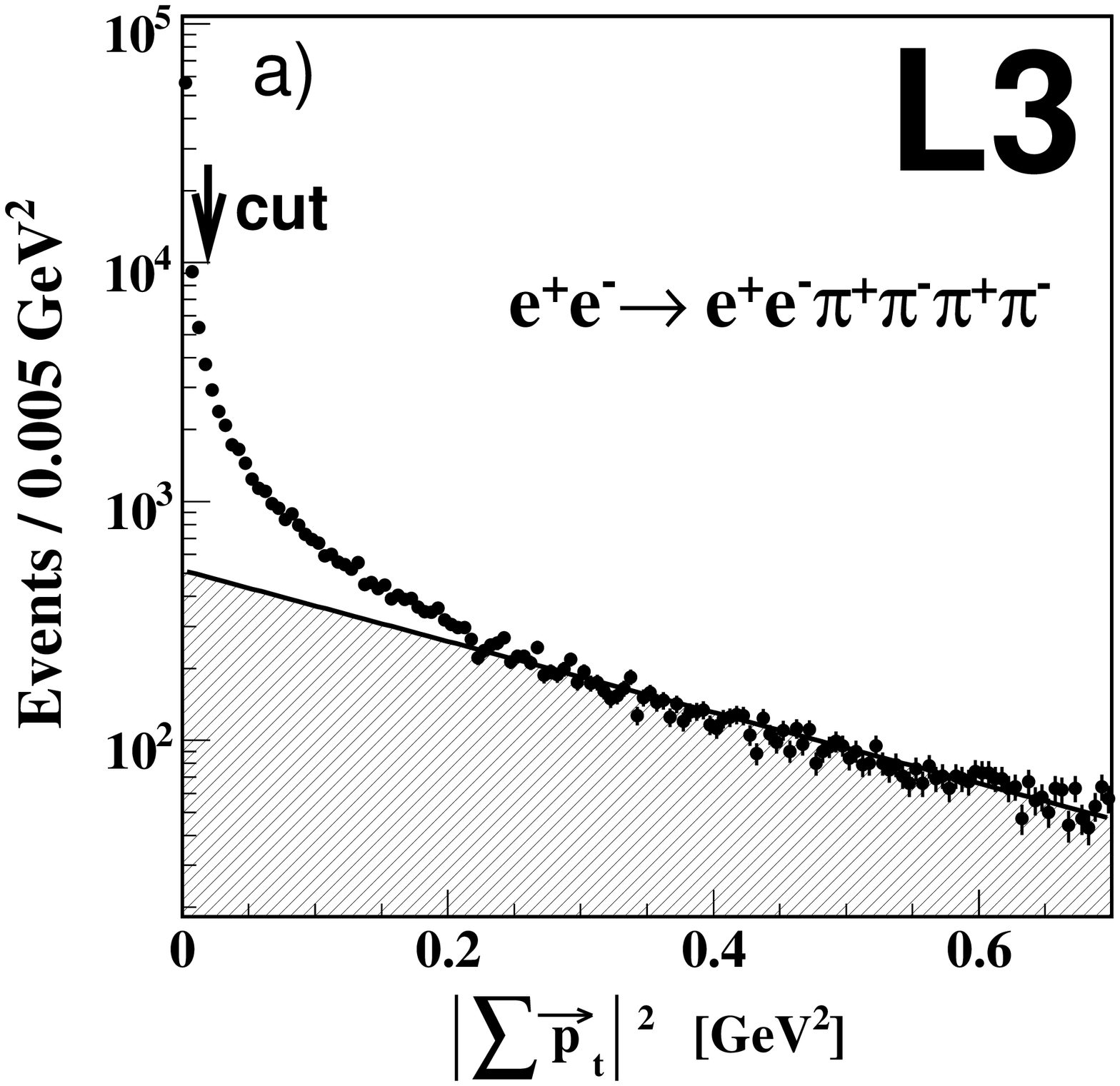&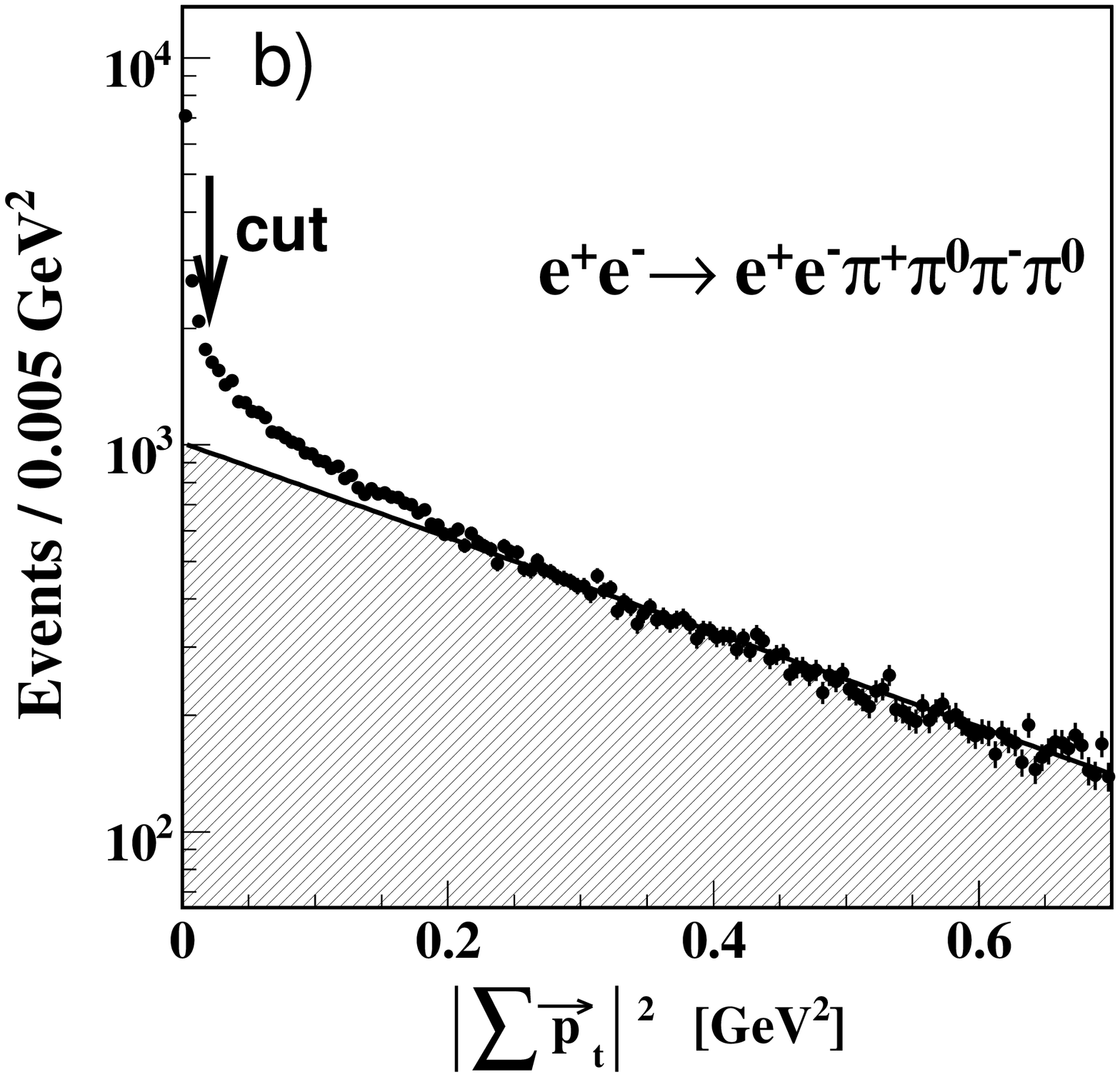\cr
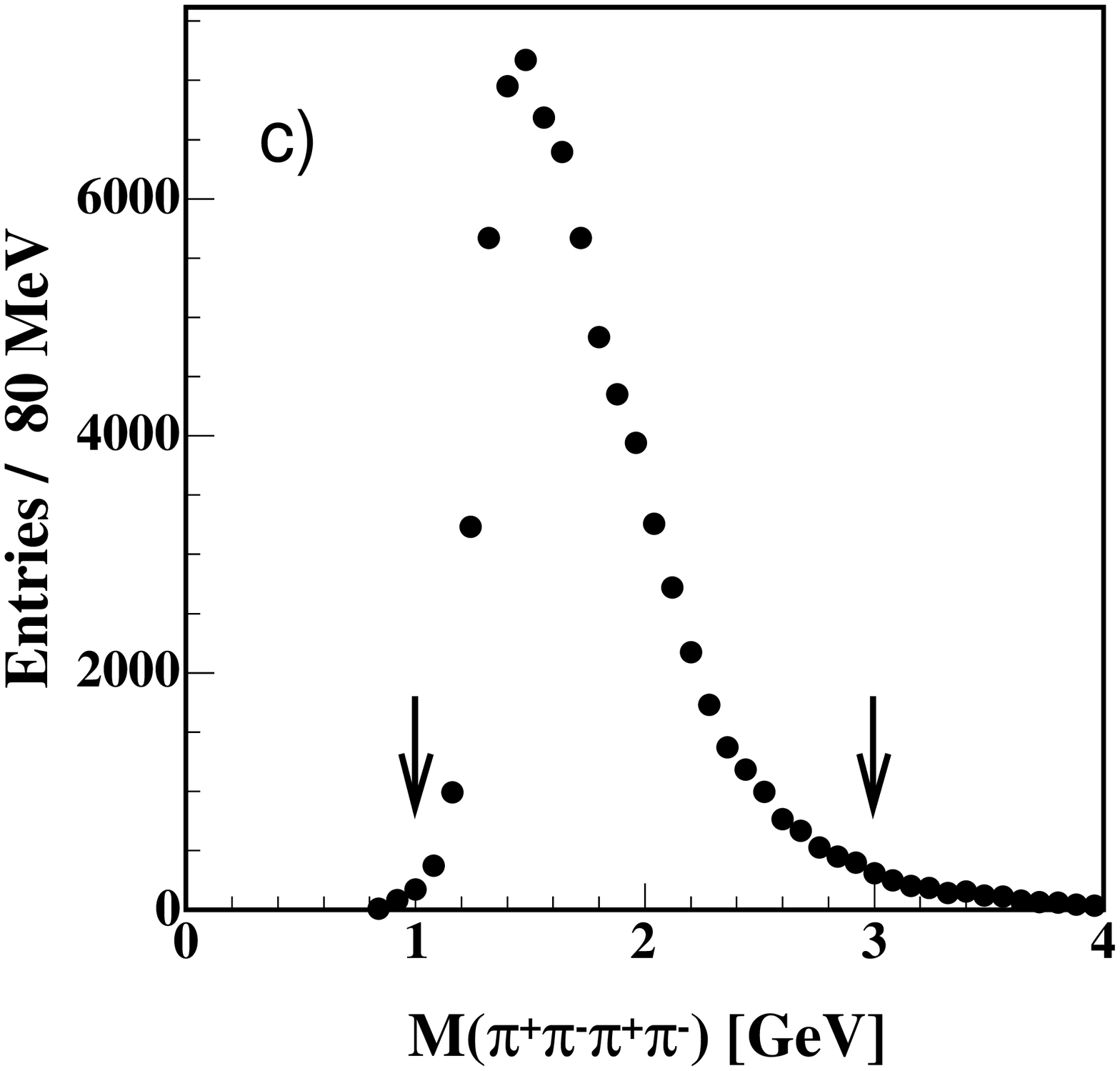&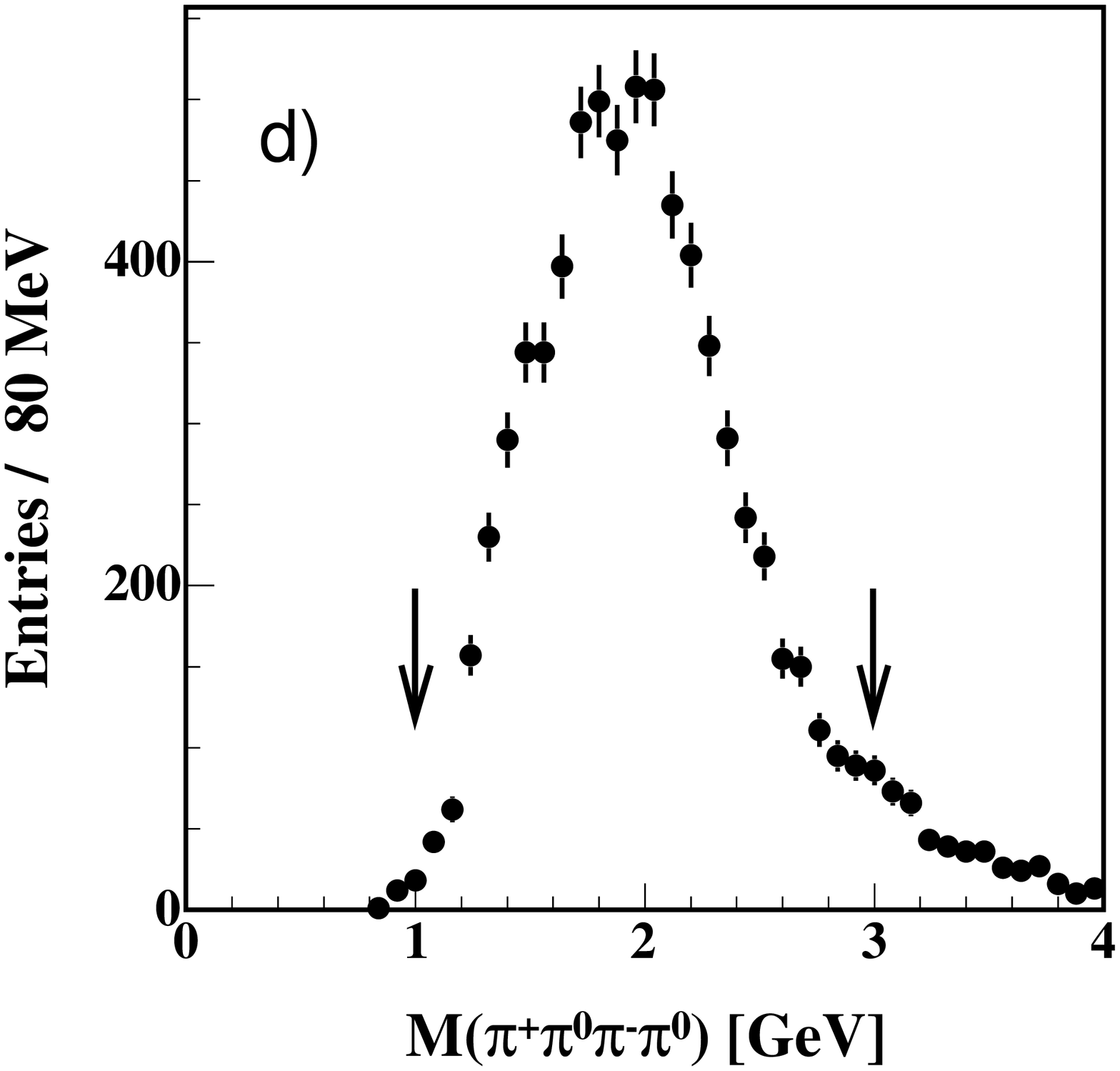\cr
}
\caption{ Distributions of 
  $|\Sigma \vec{p}_\mathrm{t}|^2 $ for  
  a)   $\eepppp$  and  b) $\eeppzppz$ events. The hatched areas
represent  the estimated non-exclusive backgrounds.
The cut values are shown by the arrows.
 Distributions of the  four-pion mass  
  for c)  $\eepppp$   and  d)  $\eeppzppz$ events.
Only events within the region indicated by the arrows are further analysed.}
\label{fig:fig1}
\end{center}
\end{figure}

To suppress the background  from non-exclusive events,  
  the overall transverse  momentum of the event, $|\Sigma\vec{p}_\mathrm{t}|^2$,
 must be less than $0.02\GeV{}^2$, 
as shown in Figures~\ref{fig:fig1}a and  \ref{fig:fig1}b.
The resulting  samples consist of \numev{} and \numevz{} events  for  the
$\eepppp$ and  $\eeppzppz$ reactions, respectively.

The distributions of the  four-pion mass, equal to $\wgg$
for exclusive events, are shown in Figures~\ref{fig:fig1}c and \ref{fig:fig1}d.
The mass resolution is estimated to be 48$\MeV$ for the $\pppp$ and 63$\MeV$
for the $\pppz$ final states. More than 90\% of the events lie in the region
$1.0\GeV \le \wgg \le 3.0\GeV$, where the spin-parity-helicity analysis 
is performed.  

The background is dominated by higher-multiplicity final states
produced in two-photon interactions which are only partially
reconstructed. The expected
contribution from annihilation events 
is negligible. As presented in Figures~\ref{fig:fig1}a and~\ref{fig:fig1}b, the
distribution of $|\Sigma\vec{p}_\mathrm{t}|^2$  for 
non-exclusive final states has an exponential form, which is 
 estimated from the data in the high $|\Sigma\vec{p}_\mathrm{t}|^2$  region,  
  $0.2\GeV^2 \le |\Sigma\vec{p}_\mathrm{t}|^2 \le 0.8  \GeV ^2$.
 Extrapolating
this exponential to the signal region, $|\Sigma{\vec p}_\mathrm{t}|^2 < 0.02 \GeV ^2$,
the   backgrounds  for the $\pppp$ and $\pppz$ final states
are estimated to be    2.5\% and   4\%, respectively.

\begin{figure}
\begin{center}
\vskip-.5cm
\includegraphics[height=0.9\textheight]{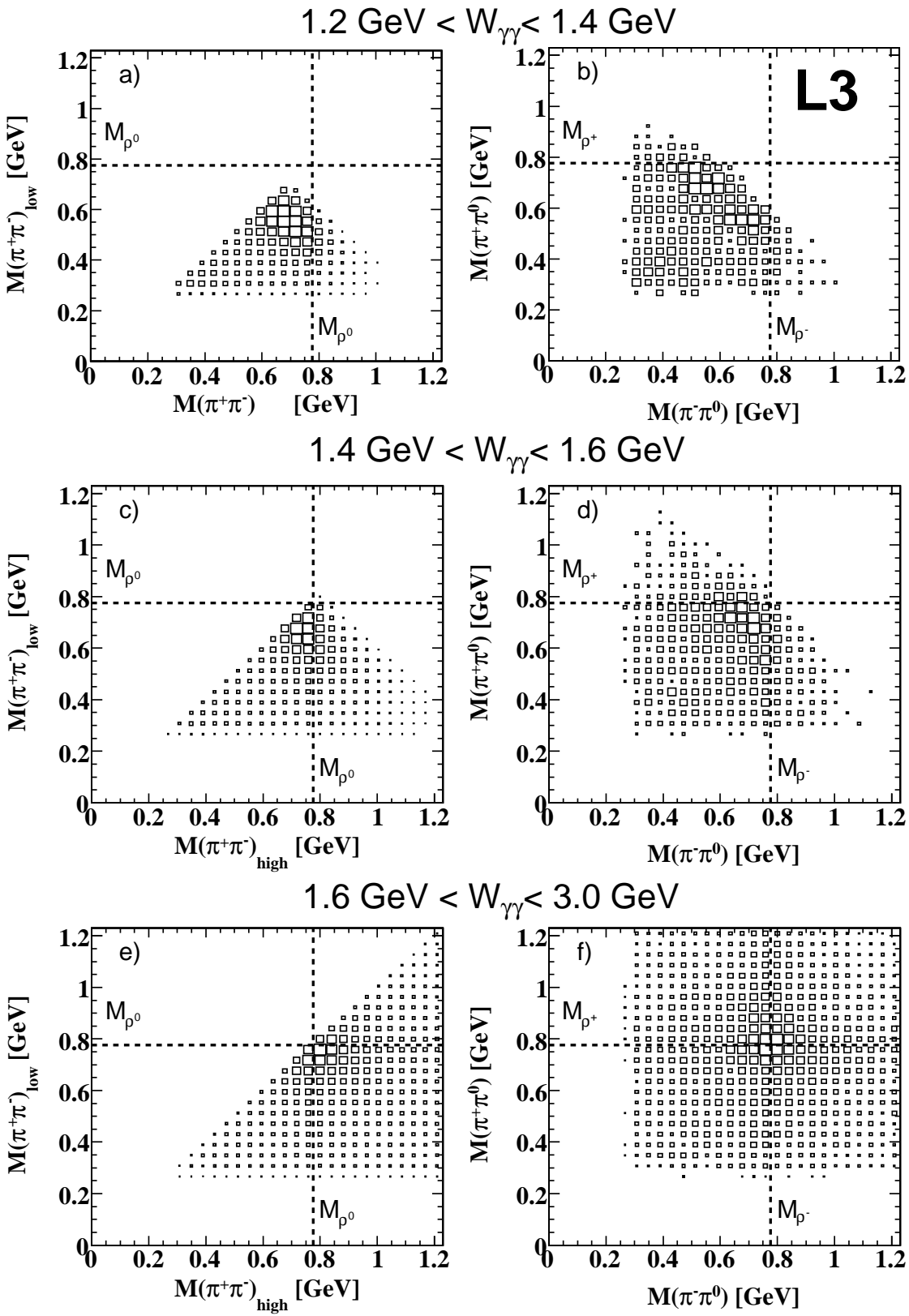}
\vspace{-0.3 cm}
\caption{Two-dimensional distributions of two-pion masses in three
different $\wgg$  regions. For a), c) and e) the  $\pip\pim$ combinations from the $\pppp$ final-state 
are shown as low-mass  \vs {} high-mass,
with two entries per event. 
In b), d) and f) the $\pi^+\piz$ \vs {} $\pi^-\piz$ combinations from the $\pppz$ final-state   
are shown, with two entries per event. The dotted lines indicate the nominal mass value of the $\rho$ meson.}
\label{fig:fig2}
\end{center}
\end{figure}

Figures \ref{fig:fig2}a, \ref{fig:fig2}c and \ref{fig:fig2}e show the
two-dimensional distributions of the masses of 
$\pi^+\pi^-$ combinations for the selected $\pppp$ events in
different $\wgg$ regions. There are two
entries per event, displayed by ordering the two masses of each
entry. Figures \ref{fig:fig2}b, \ref{fig:fig2}d and \ref{fig:fig2}f
show the  $\pi^+\pi^0$ and $\pi^-\pi^0$ mass combinations for
the $\pppz$ channel with two entries per event.

The two-pion mass resolution is estimated from 
Monte Carlo simulation to be 25$\MeV$ for both the $\pi^+\pi^-$ and
$\pi^\pm\pi^0$ cases.
The $\pi^+\pi^-$ and $\pi^\pm\pi^0$
combinations shown in Figure~\ref{fig:fig2} present clear evidence of
$\rho\rho$
production. For $\wgg<1.6\GeV$, the $\rho$
signal is distorted by threshold effects.
As $\wgg$ increases, the $\rho$ signal shifts to its nominal mass 
value, shown by the dotted lines in the figure.

%
%
\section{Spin-Parity-Helicity Analysis}

Following  the model proposed by the TASSO Collaboration~\cite{TASSO1},
we consider  $\rho\rho$
production in different spin-parity and helicity states $(J^P,J_z)$,
together with an isotropic production of four pions, denoted as ``$4\pi$".
 All states are assumed to be produced incoherently, and
therefore no interference effects between the
final states are taken into account.
However, since states of different  spin-parity and helicity 
are orthogonal, all interference terms vanish  on integrating over the angular 
phase space.
Isotropic $\rho\pi\pi$ production, included in previous 
analyses~\cite{ARGUS,ARGUS0}, corresponds to an unphysical state since $C$-parity requires the angular
momentum between the two pions to be odd.
We have verified that this state is not essential to reproduce 
the data. The  $\rho\pi\pi$ events, if neglected,  are absorbed by 
  the $4\pi$  background.
 
The analysis is performed in $\wgg$ intervals of 100$\MeV$ for $\ggpppp$
and 200$\MeV$ for $\ggppzppz$. 
 As pions are bosons, the amplitudes which describe the
process must be symmetric under  interchange of two pions with the same
charge and are:
$$ g_{J^P\,J_z} = B_\rho(m_{\rho_1})
B_\rho(m_{\rho_2})\Psi_{J^PJ_zLS}(\rho_1,\rho_2) +
{\rm permutations},$$ 
 $$ \mathrm{and} \;\;\; g_{4\pi}  =  1,$$
where $m_\rho$ indicates the mass of the two-pion system and
 $B_\rho(m_{\rho})$ is the relativistic 
Breit-Wigner amplitude for the $\rho$
meson~\cite{Jack}. 
The angular term $\Psi_{J^PJ_zLS}(\rho_1,\rho_2)$  
describes the rotational properties of the $\rho\rho$ state with  
spin-parity $J^P$ and helicity $J_z$. It is constructed by combining the spins 
of the two $\rho$ mesons, $\vec{S}=\vec{s}_1+\vec{s}_2$, with $z$ projection $ M_s = m_1 +m_2$
and then  adding this to the $\rho\rho$ 
orbital angular momentum, $\vec{L}$, with $z$ projection $ M$, to obtain the state with 
total angular momentum $\vec{J}$ and $z$ projection $ J_z = M_s +M$:
$$ 
\Psi_{J^PJ_zLS}=\sum_{M,m_1} C^{J^PJ_z}_{LMSM_s}
C^{SM_s}_{s_1m_1s_2m_2} Y_{L 
M}(\xi_1)Y_{s_1m_1}(\xi_2)Y_{s_2m_2}(\xi_3)
$$
where $C^{JM}_{l_1m_1l_2m_2}$ are  the Clebsch-Gordan coefficients, 
$Y_{lm}(\xi_i)$ are the spherical harmonics and
$\xi_1=(\vartheta_\rho,\varphi_\rho)$, 
$\xi_2=(\vartheta_{\pi^+_1},\varphi_{\pi^+_1})$  
and
$\xi_3=(\vartheta_{\pi^+_3},\varphi_{\pi^+_3})$, with $\vartheta_\rho$ and
$\varphi_\rho$  being
 the polar and azimuthal angles of a $\rho$ meson in the 
two-photon helicity system.  
The $z$ axis is chosen parallel to the beam 
direction, which to a good approximation is parallel to the $\gaga$ helicity 
axis. The angles $\vartheta_{\pi^+_1}$ and  $\varphi_{\pi^+_1}$
are the polar and azimuthal angles of the positive pions in the
centre-of-mass of  
the first  $\rz$ meson, with the  $z$ axis parallel to the beam axis, the
angles $\vartheta_{\pi^+_3}$ and  $\varphi_{\pi^+_3}$ correspond to
the second $\rz$ meson;
for a $\rho^-$ meson,
$\xi_3=(\vartheta_{\pi^-_3},\varphi_{\pi^-_3})$.
The indices from 1 to 4 refer to the four pions using the
convention: $\pip_1 \pim_2 \pip_3 \pim_4$ or $\pip_1 \piz_2 \pim_3 \piz_4$.
Since the analysis is performed close to threshold, the orbital
angular momenta are restricted to $L=0,1$. The allowed
spin-parity-helicity final states of the 
$\rho\rho$ system in quasi-real  two-photon reactions are then: 
$(J^P,J_z)=0^+$, $0^-$, $(2^+,0)$, $(2^+,\pm2)$ and $(2^-,0)$, with the 
total spin of the 
$\rho\rho$ meson system  $S=1$ or $S=2$. States with helicity one
are forbidden by  helicity 
conservation and spin-one states by the Landau-Yang theorem \cite{Landau}.

 A maximum-likelihood fit to the data is used in each $\wgg$ bin to
determine the contributions of the four amplitudes: $4\pi$,
 $0^+$, $0^-$ and $(2^+,2)$.
The remaining spin-parity states are not considered as they have a negligible
contribution if included in the fit.

\section{Cross Section}

\begin{figure}[htbp]
\begin{center}
\halign{&\includegraphics[width=.45\textwidth]{#}\cr
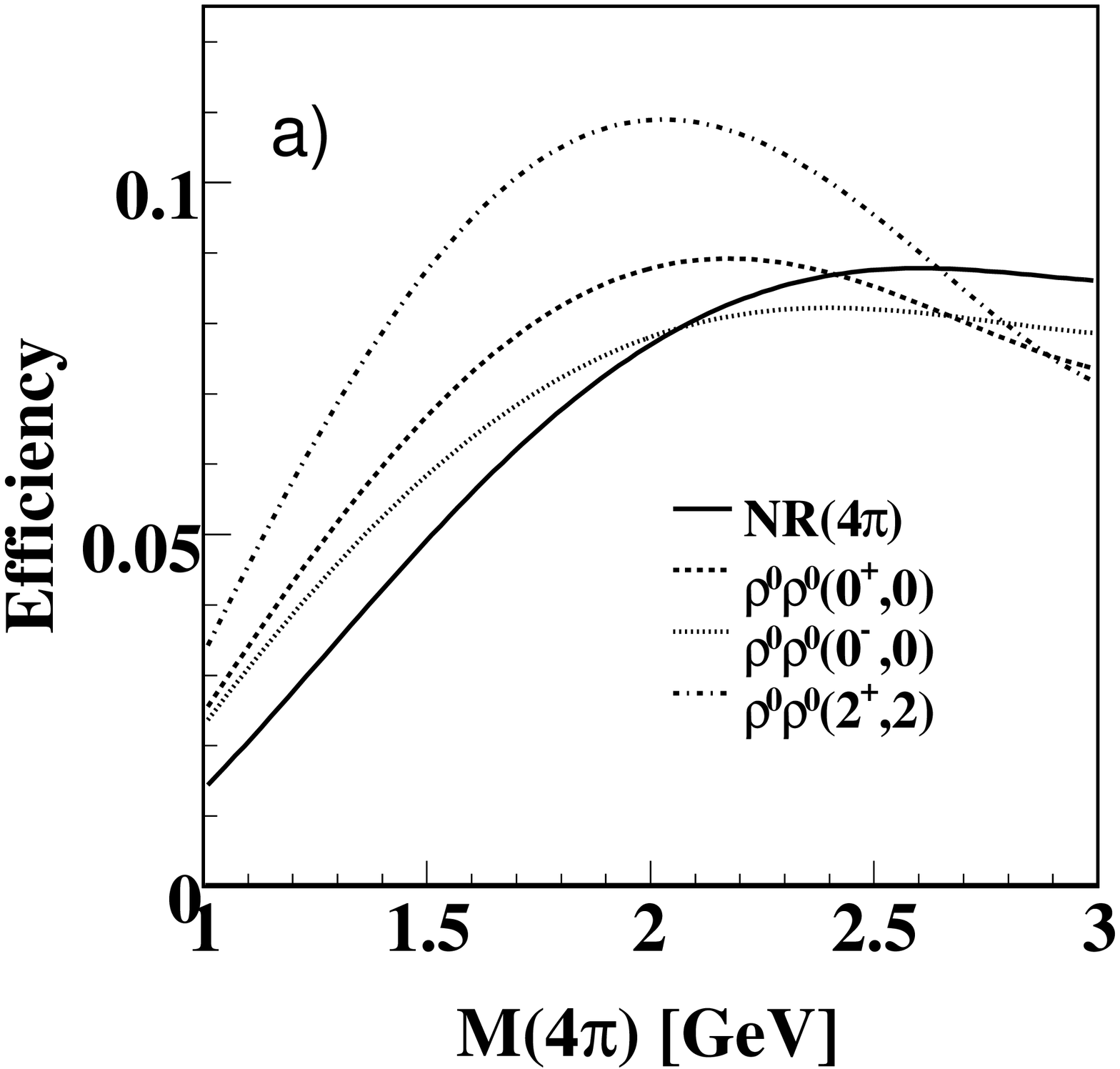&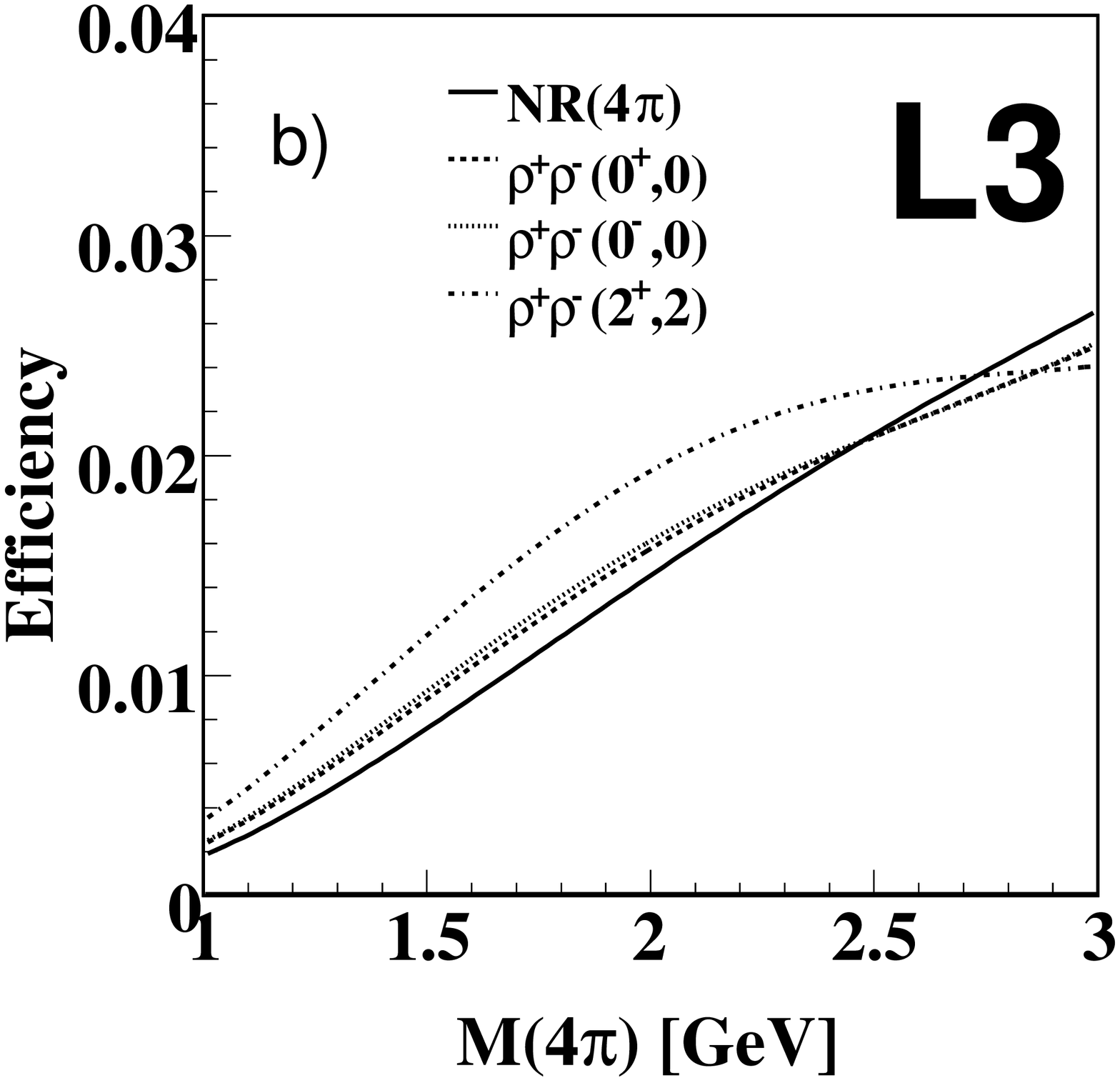\cr
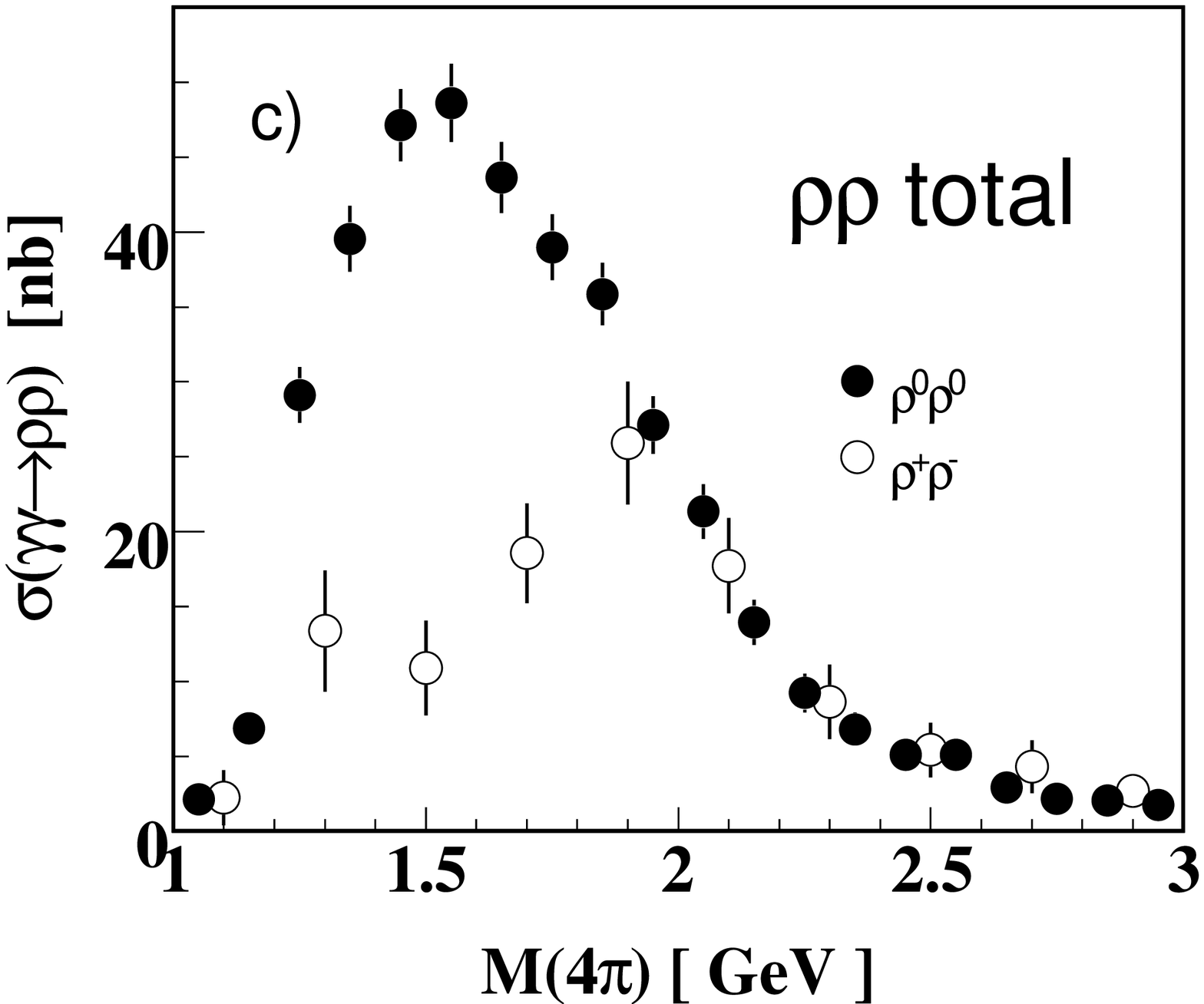&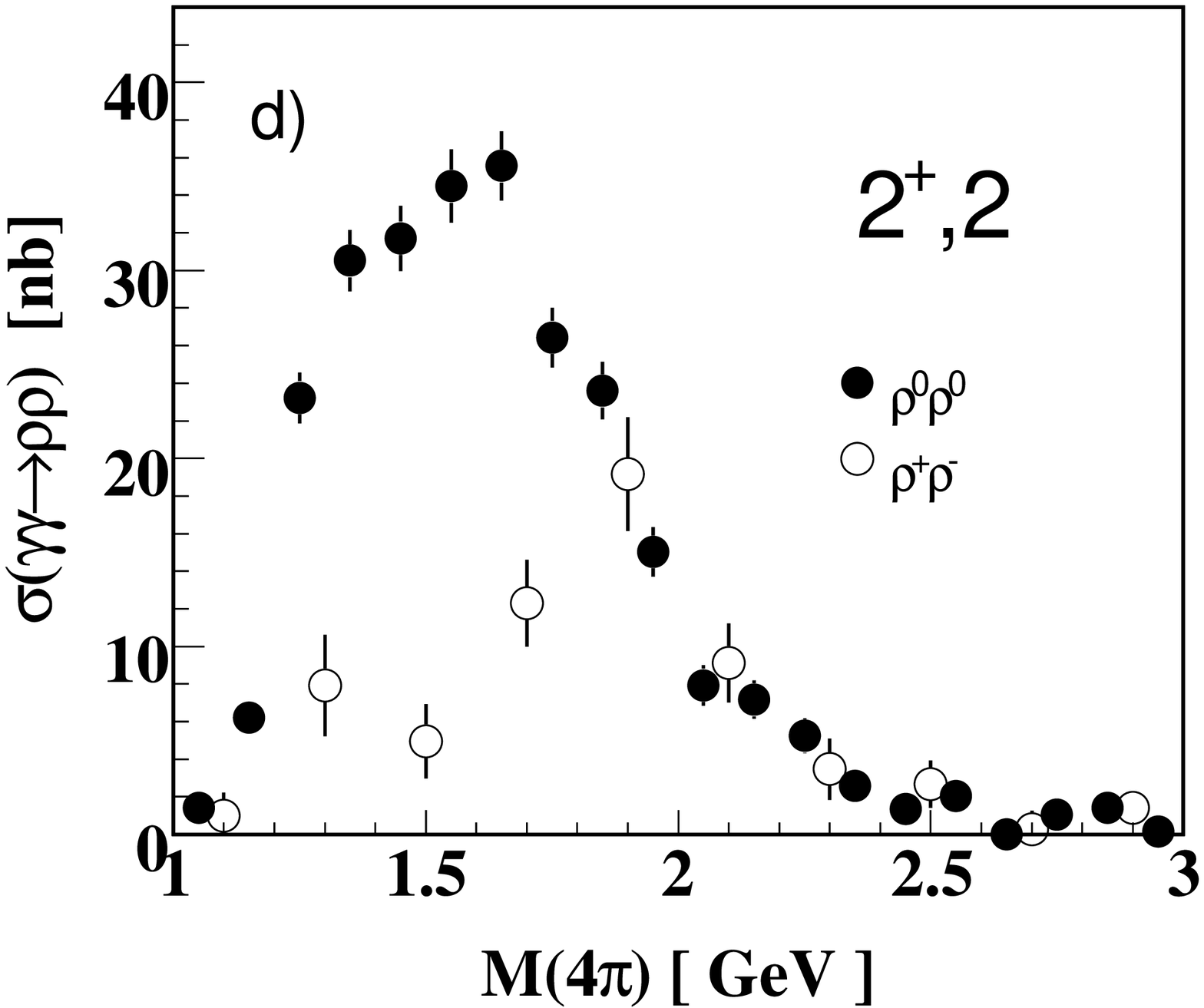\cr
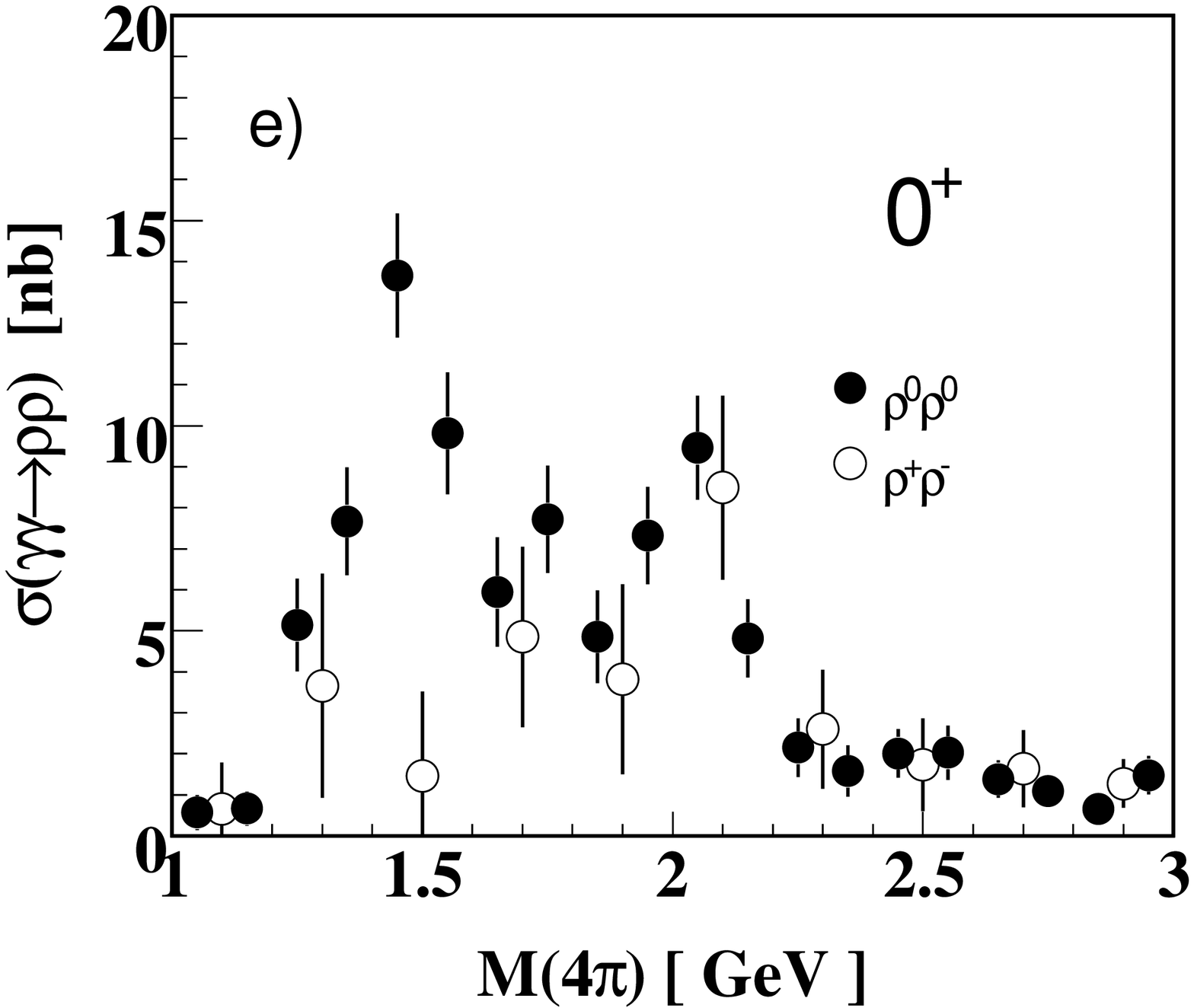&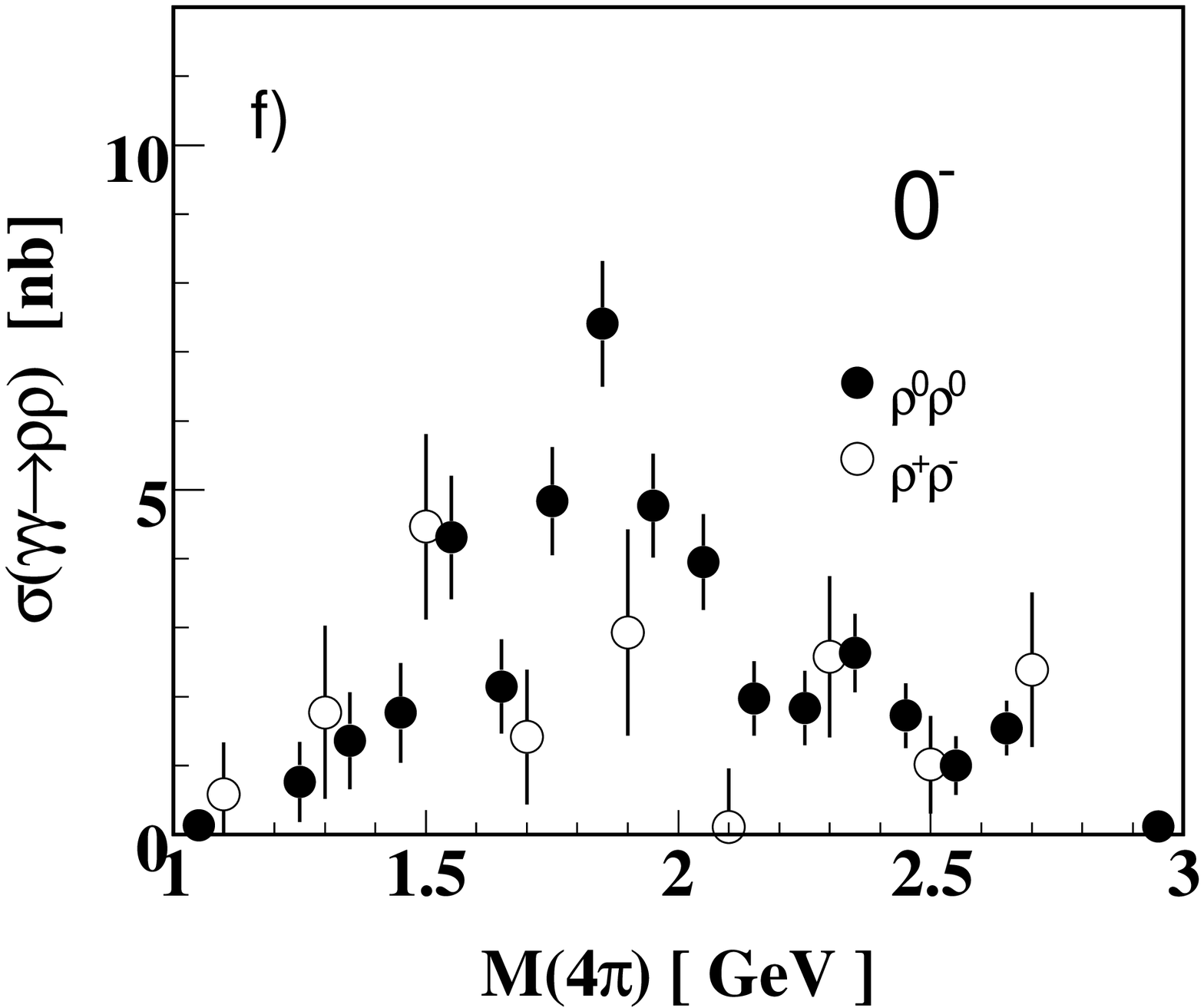\cr
}
\vspace{-0.3 cm }
\caption{
Selection efficiencies for the different contributions to the a)
$\pppp$ and b) $\ppzppz$ final states.
Measured  cross sections for the $\eepppp$ and $\eeppzppz$ processes: 
c) the total  $\ggrzrz$ and $\ggrprm$ cross sections, d) $(2^+,2)$, e) $0^+$, f) $0^-$  contributions.
 The error bars show the statistical 
uncertainties.}
\label{fig:fig9}
\end{center}
\end{figure}

The cross section for the process {\it k}, with fraction $\lambda_k$
determined from the fit,  averaged over the $\wgg$
bin with $N$ events, is  

$$\sigma_{\gaga\ra k}=\frac{N\lambda_k}{
  {\mathcal{L}}_{\epem}\;\varepsilon_k(\wgg)\;\varepsilon_\mathrm{trg}(\wgg)\;
\int \d\mathcal{L}_{\gaga}} \;\;,$$
where $\int \d{\mathcal{L}}_{\gaga} $ is the   
two-photon luminosity function integrated over the $\wgg $ bin, $\varepsilon_k$ is the  
selection efficiency  and $\varepsilon_\mathrm{trg}$ is the trigger efficiency.
The selection efficiencies depend on $\wgg $ as well as  on the particular wave. They are 
computed by Monte Carlo simulation, re-weighting the events with 
the amplitudes $|g_k|^2$.  The efficiencies for the $4\pi$ process are
listed in  Tables~\ref{tab:xsec0} and~\ref{tab:xsec1}. Similar
efficiencies are found for the other processes.
The trigger efficiency is studied 
by comparing the response of the  two charged-track triggers. The higher-level
trigger  efficiencies are determined using  prescaled   
events. The total trigger  efficiency  
 is given in Tables~\ref{tab:xsec0} and~\ref{tab:xsec1}. The overall efficiencies for the 
$\pppp$ and $\ppzppz$ final states are shown in Figures~\ref{fig:fig9}a and~\ref{fig:fig9}b.

%
The cross sections derived from the fit are presented
in Tables~\ref{tab:xsec0} and~\ref{tab:xsec1}. 
 Figures~\ref{fig:fig9}c$-$f compare the total cross sections 
and the contributions of the individual
waves to the 
$\ggrzrz$ and $\ggrprm$ processes
 as a function of the four-pion mass.  The $4\pi$ background,
which in this analysis represents all states which do not correspond
to the $\rho\rho$ production hypothesis, is similar in both
channels. It grows from threshold to a value of $20-30 \; \mathrm{nb}  $  around $2 \GeV $
and decreases  toward $3 \GeV $.
In the $\pppp$ channel, the $\rzrz$
production has a high cross section, with a maximum of about 50 nb at
$1.6 \GeV $. It is dominated by the $(2^+,2)$
 state, which has a cross section peak-value 
of about 35 nb.
The $\pppz$ channel, exhibits also a significant $(2^+,2)$ contribution, but
only for $1.6 \le \wgg \le  2.2\GeV $. 
 Above $1.9 \GeV$ the   $\ggrzrz$ and $\ggrprm$ cross sections are equal
within the experimental uncertainties and fall rapidily with increasing $\wgg$.
In the $0^+$ wave, a clear peak is observed in the $\rzrz$ channel at $\wgg \simeq 1.4 \GeV$,
perhaps indicative of an $s$-channel resonance effect, it is absent in the  $\rprm$ channel. The $\ggrprm$ cross
section peaks near   $2 \GeV$ in both $0^+$ and $(2^+,2)$ waves, while a similar peak is seen for 
the $\rzrz$ in the $0^+$ wave only. The same qualitative features 
were observed by the ARGUS Collaboration~\cite{ARGUS,ARGUS1}, which 
however found a higher peak cross section of $\simeq 50 $ nb for the 
($2^+ , 2$) wave in $\rzrz$. Taking into account the  larger
experimental uncertainties on the ARGUS data, a peak is also seen in the  $0^+$
wave at $\wgg \simeq 1.4 \GeV$. However, at higher mass values, $\wgg > 2 \GeV $,
only the much higher statistics of the present experiment are able to provide 
cross section measurements, so no comparison is possible.

\begin{figure}[htbp]
\begin{center}
\halign{&\includegraphics[height=0.31\textheight,width=.5\textwidth]{#}\cr
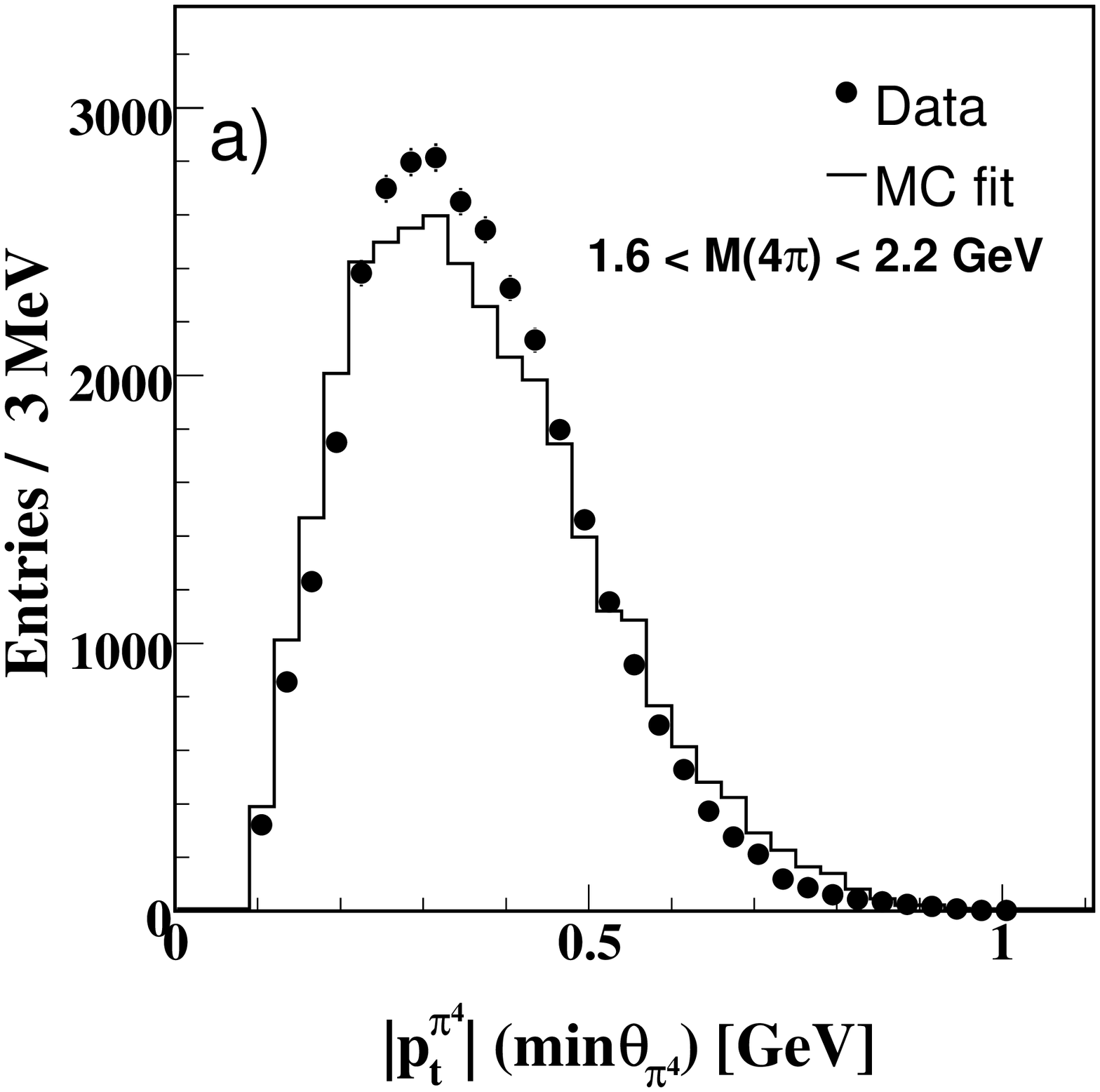&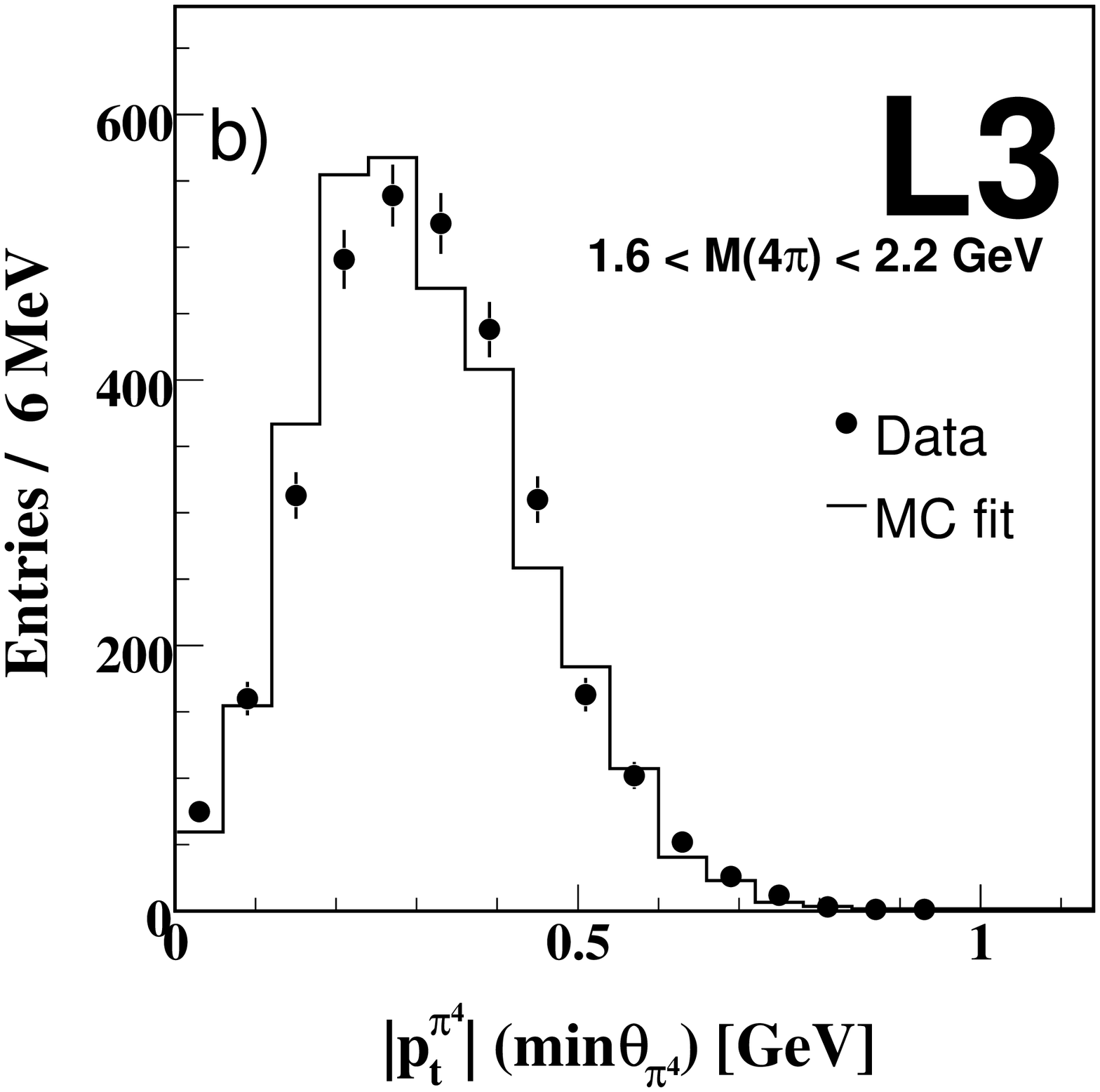\cr
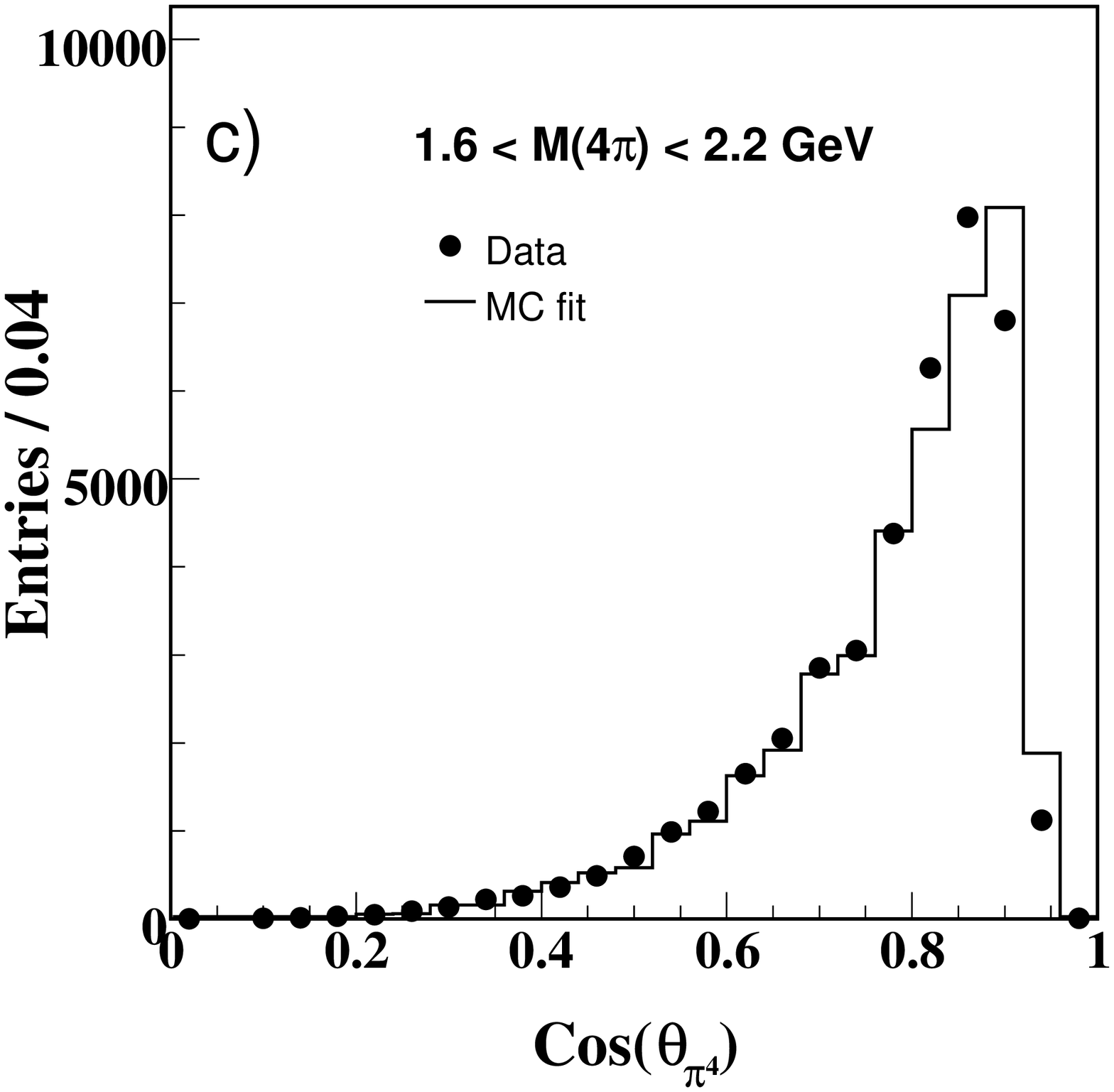&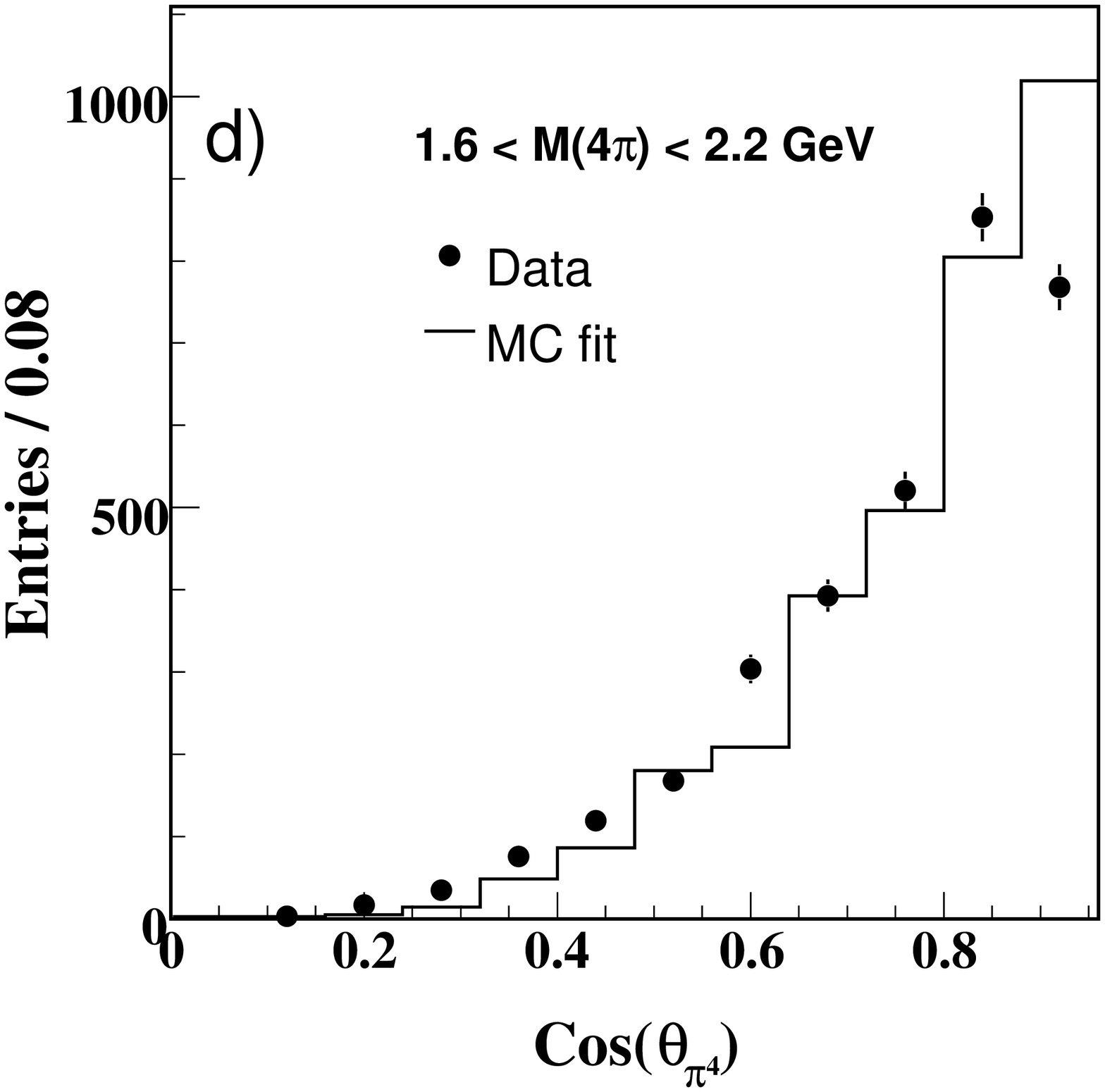\cr
}
\vspace{-0.3 cm }
\caption{Comparison of the Monte Carlo simulation normalised to the
  fit results to the data: a) and b) transverse momentum of the charged
  or neutral pion closest to the beam line, respecively; c) and d) cosine of the polar angle of
the charged or neutral pion closest to the beam line,
  respectively. The statistical uncertainty on the Monte Carlo
  distributions -not shown- is comparable to that of the data.}

\label{fig:fignew}
\end{center}
\end{figure}

\begin{figure}[p]
\begin{center}
\offinterlineskip\tabskip0pt
\halign{&\includegraphics[height=0.37\textheight,width=.5\textwidth]{#}\cr
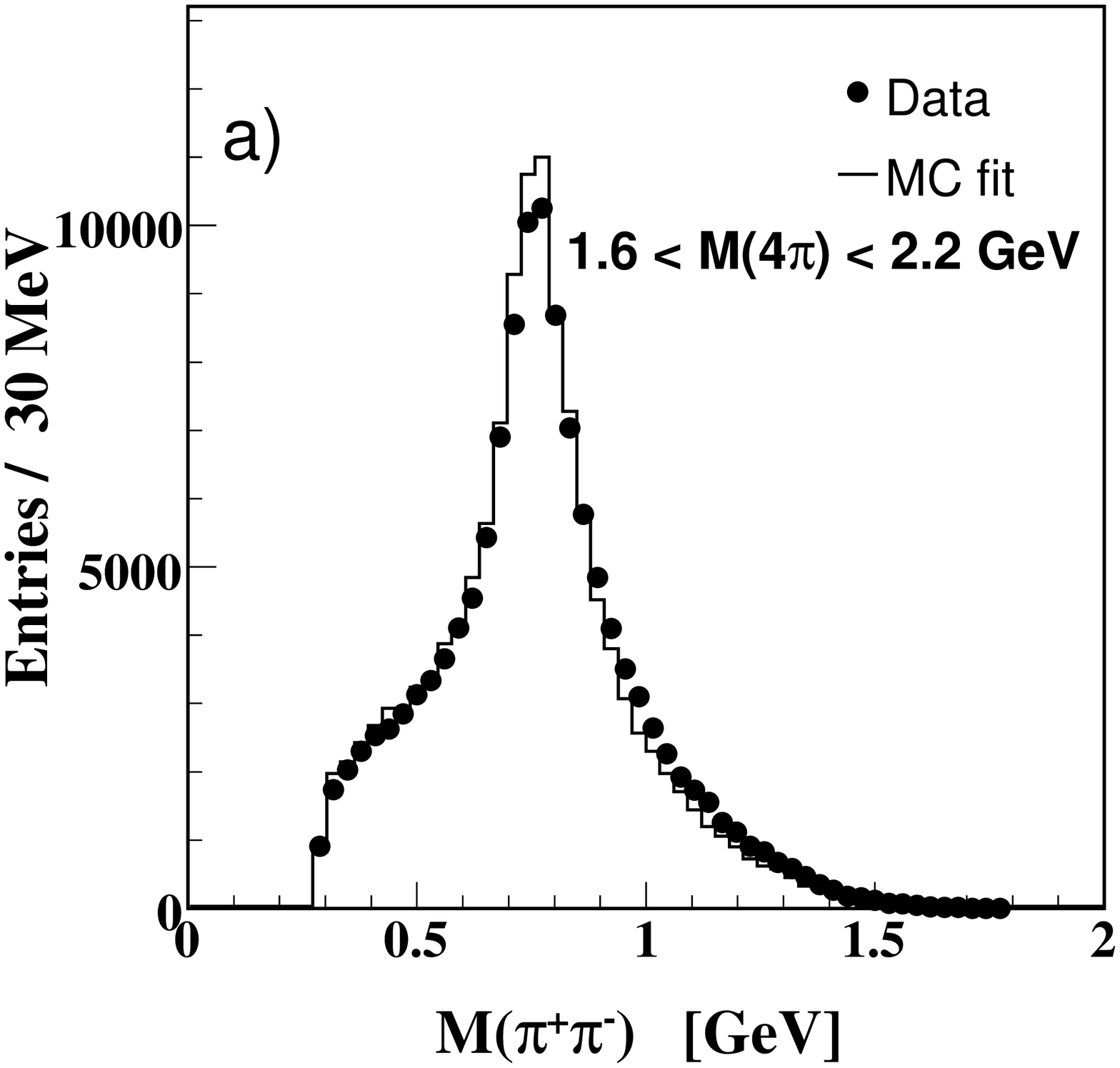&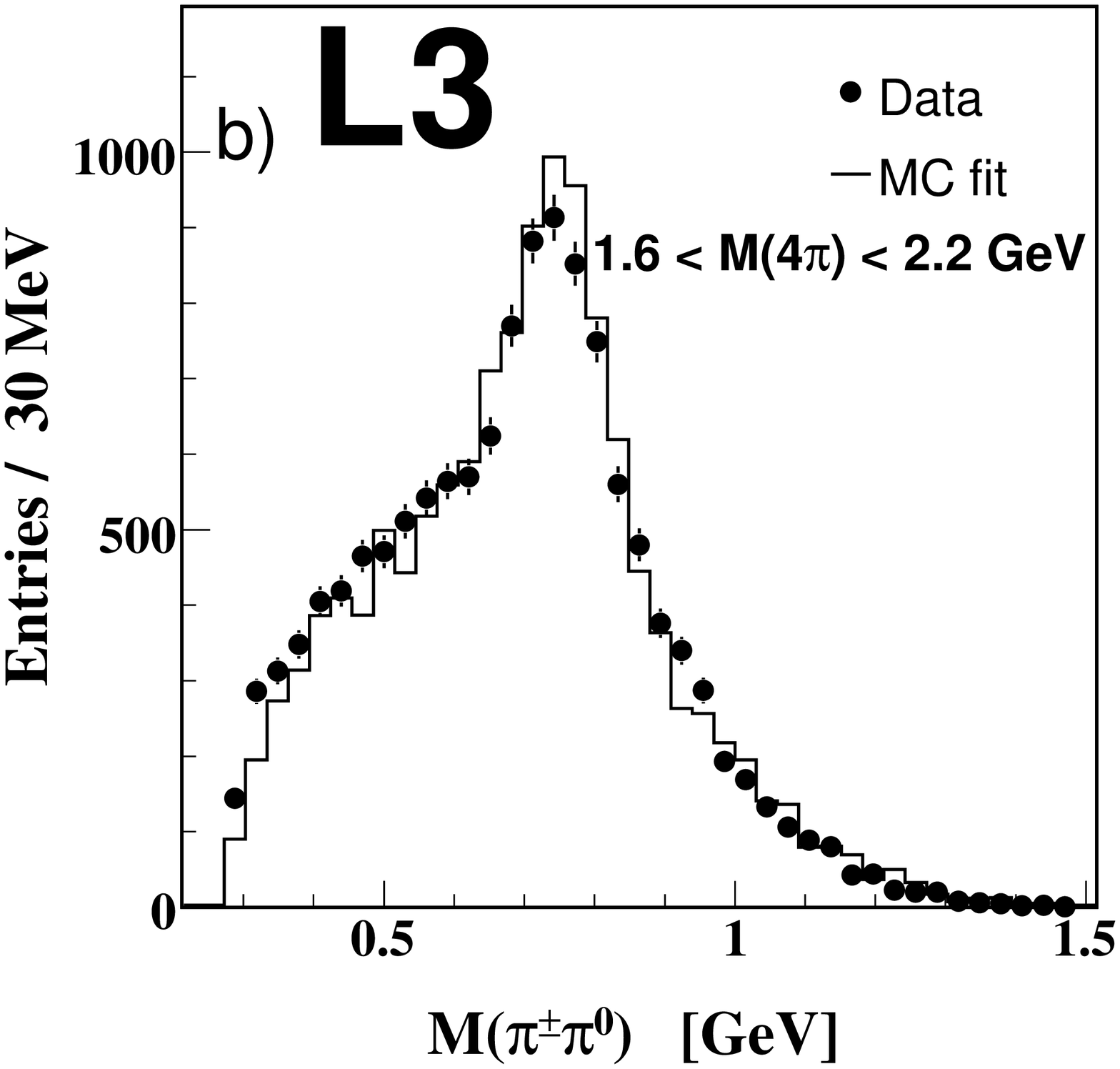\cr
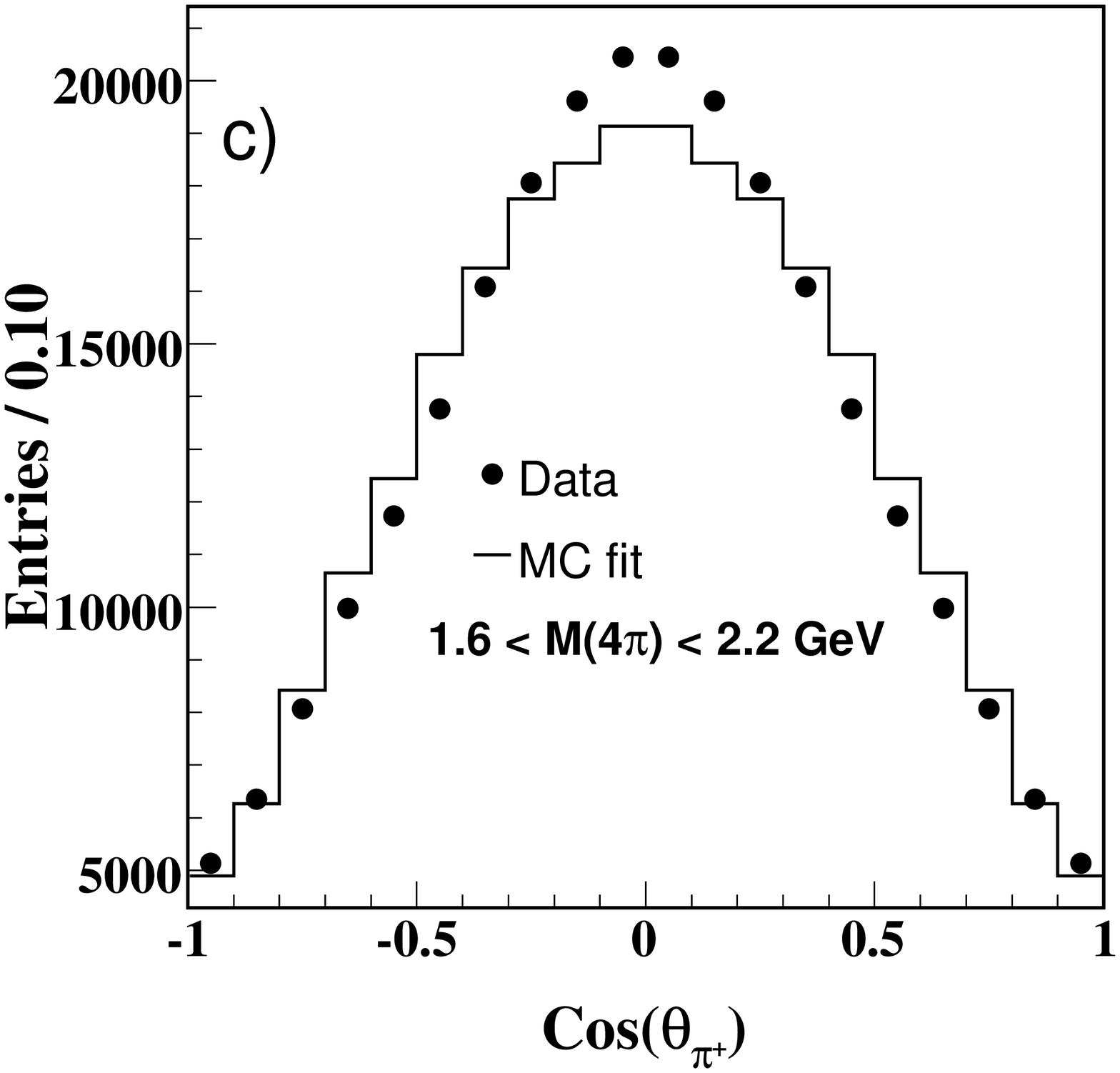&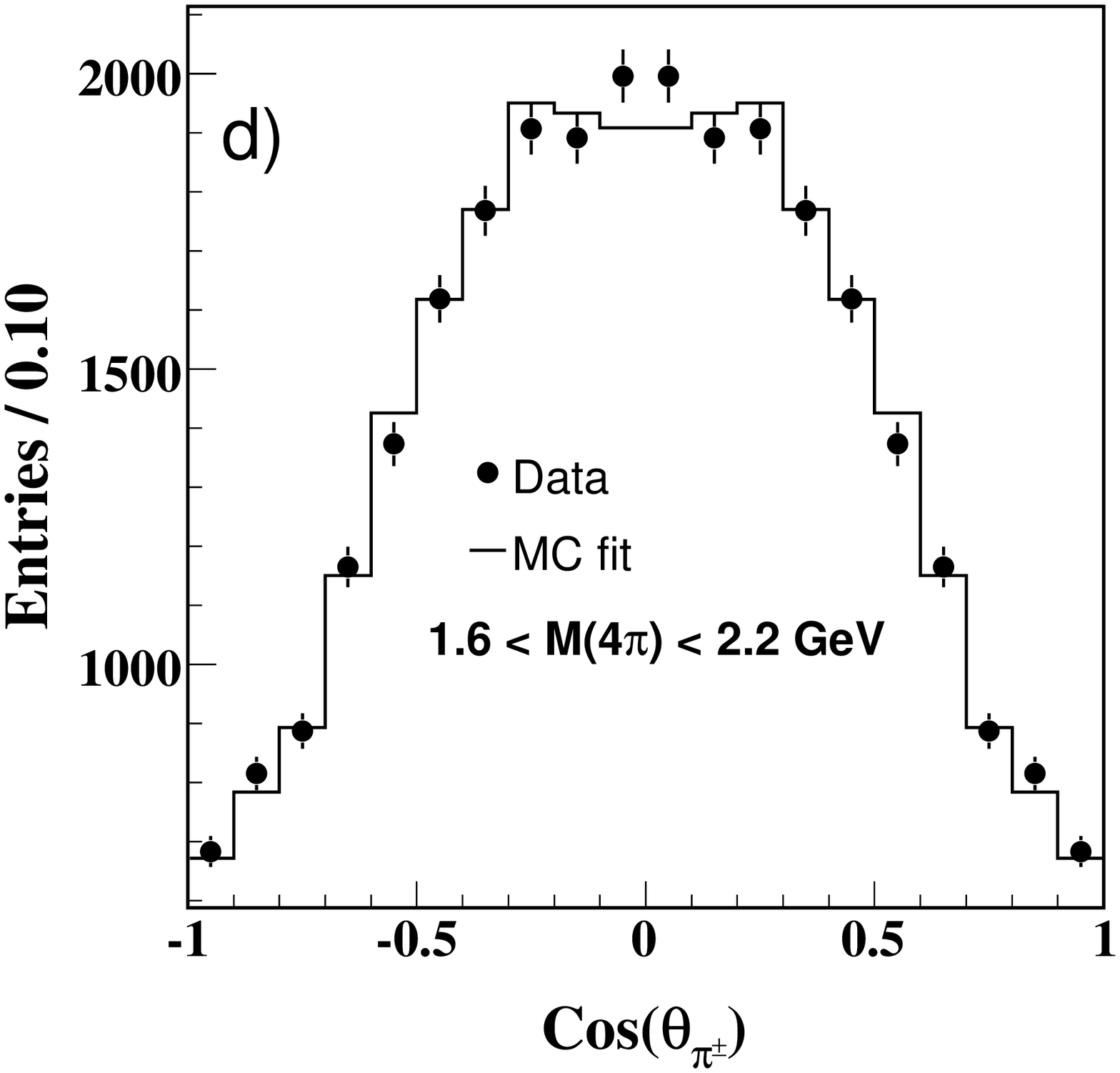\cr
}
\caption{Comparison of the Monte Carlo simulation normalised to the fit results to the data: a) two-pion
opposite-sign mass combinations for $\ggpppp$ (four entries per event)
b) two-pion charged mass combinations for $\ggppzppz$ (four entries per
event) c) $\cos\theta_{\pi^+}$, where $\theta_{\pi^+}$ is the
production angle with respect to the beam axis in the unlike-sign  two-pion
centre-of-mass system for $\ggpppp$ (four entries per
event) and d) $\cos\theta_{\pi^\pm}$, where $\theta_{\pi^\pm}$
corresponds to the $\theta_{\pi^+}$ angle for the $\pppz$ system (four
entries per event).
The error bars show the statistical uncertainties.}
\label{fig:fig10}
\end{center}
\end{figure}

To evaluate the quality of the fit and of the detector modeling we
compare several distributions of the data with a Monte Carlo
simulation 
normalized to the fit results. Figure~\ref{fig:fignew} shows the
distributions of $p_{\rm t}$ and of the cosine of the polar angle of
the charged or neutral pion closest to the beam line.
The two-pion mass combinations, $\pi^+\pi^-$ and $\pi^\pm\pi^0$,
and the production angles  of the pions
in the two-pion centre-of-mass system with respect to the beam
direction (Adair angle)
are plotted in Figure~\ref{fig:fig10}.
Four entries per event are considered and the data are plotted
before acceptance corrections.
 The agreement with the Monte Carlo simulation is adequate, considering
the simplicity of the model and the high statistics of the data sample.
The angular distributions in Figures~\ref{fig:fig10}c and  \ref{fig:fig10}d
are of the general form $\sin ^2 \theta$, indicating a dominantly transverse
polarisation for the produced $\rho$.

%
%
\section{Background Estimation and Systematics}
The fraction  of non-exclusive background in the $\roro$ sample is
derived by 
performing a spin-parity-helicity analysis of the background data
sample, defined as  the region  $0.2\GeV^2< |\Sigma\vec{p}_\mathrm{t}|^2 < 0.8 \GeV ^2$. We find 
that less than 30\% of these events are classified as
$\rho\rho$. The background contribution in the $\roro$ sample is then
of the order of 1\%.

Systematic uncertainties on the $\roro$ cross sections  
 are due to   selection criteria,  fitting procedures and trigger efficiencies.
Uncertainties from the selection procedure are estimated by varying the
cuts on the quality  of the event and on the track
definition. They vary between $3 \% $ and $10\% $ for the $\rzrz$ channel
and between  $10\% $ and $24\% $ for the  $\rprm$ channel, depending on  $\wgg $.
 Uncertainties on the model and  the fitting procedure are
estimated by  neglecting 
in turn the   $0^+$ and $0^-$ waves and including the ($2^+,0$) and ($2^-,0$)  waves   
in the fit. Small effects from the inclusion of additional spin-parity states are also
considered. In the high mass region, $\wgg > 2 \GeV$, the contribution
of other channels and higher-angular momentum states may become important.
It was found that such effects could be modelled by including a contribution 
from the isotropic $\rho\pi\pi$ production.
 In total, these   uncertainties for 
 the $\rzrz$ and $\rprm$  channels  amount to a maximum of $10\% $ and $19\% $, respectively, 
for $\wgg <2 \GeV $ and to a maximum of $60\% $ in the region
$2\GeV <\wgg <3 \GeV $.
 Uncertainties on the determination of the trigger efficiencies
are of a statistical nature and affect mainly  the $\pppz$ channel,
where they vary between $2\% $ and $6\% $. They are
below 1\% for the $\pppp$ channel. Uncertainties on the background
level are below 1\% for both channels.

\section{Discussion}
\begin{figure}[htbp]
\begin{center}
\includegraphics[width=\textwidth]{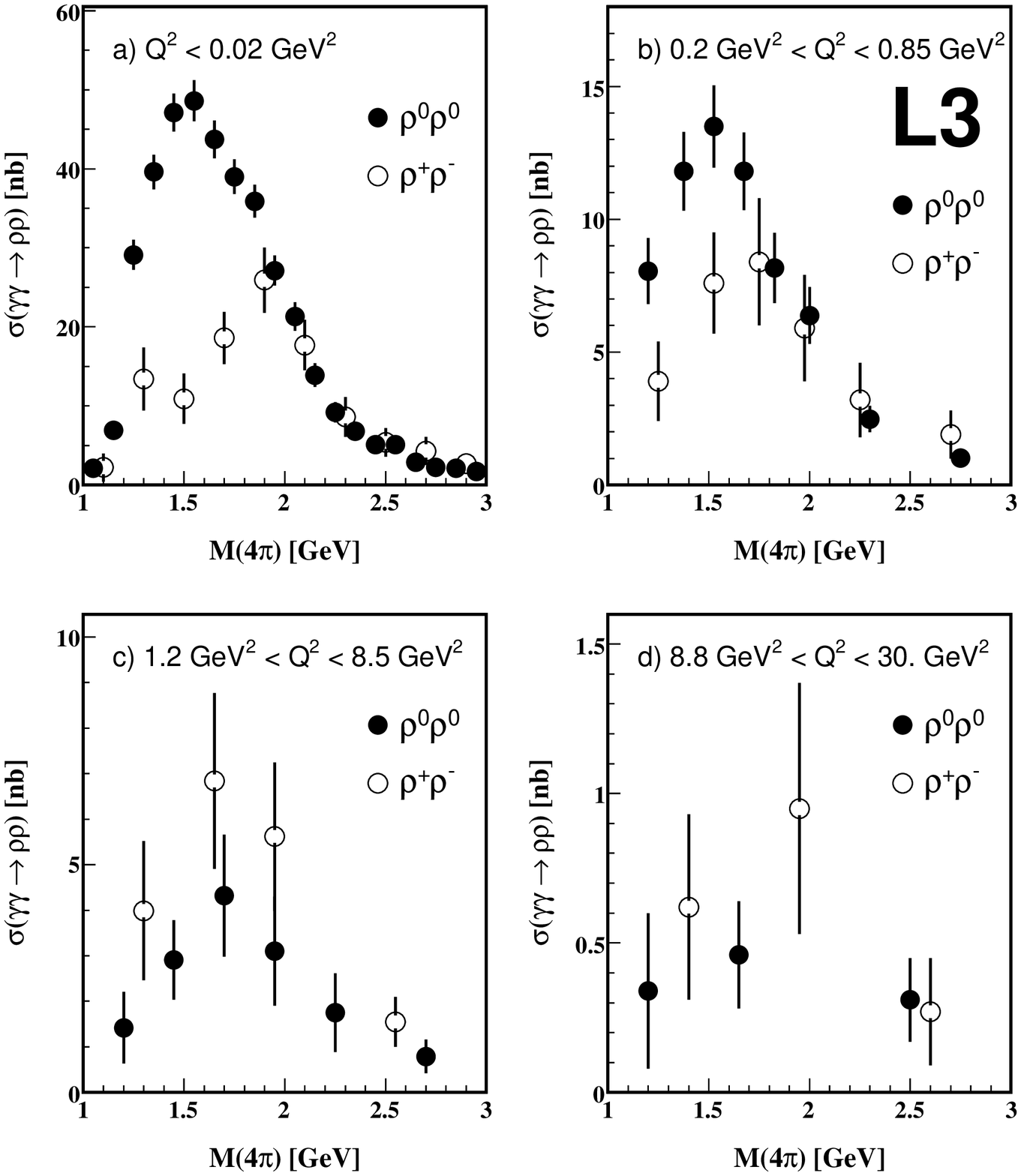}
\vspace{-0.3 cm }
\caption{ The $\ggrzrz$ and $\ggrprm$ cross sections as a function of the 
four-pion mass a) as obtained in the present analysis at $ Q^2 \le 0.02 \GeV ^2$   compared to 
previous L3 results obtained at 
 b) $ 0.20 \GeV^2 \le Q^2 \le 0.85 \GeV^2 $\protect\cite{vsat00,vsatch},  
 c)  $ 1.2 \GeV^2 \le Q^2 \le 8.5 \GeV^2  $\protect\cite{high00,highch}  and 
 d)  $  8.8 \GeV^2 \le Q^2 \le 30  \GeV^2  $\protect\cite{high00,highch}.
 The error bars show the statistical uncertainties.}
\label{fig:fig11}
\end{center}
\end{figure}

\begin{figure}[htbp]
\begin{center}
\includegraphics[height=.9\textheight]{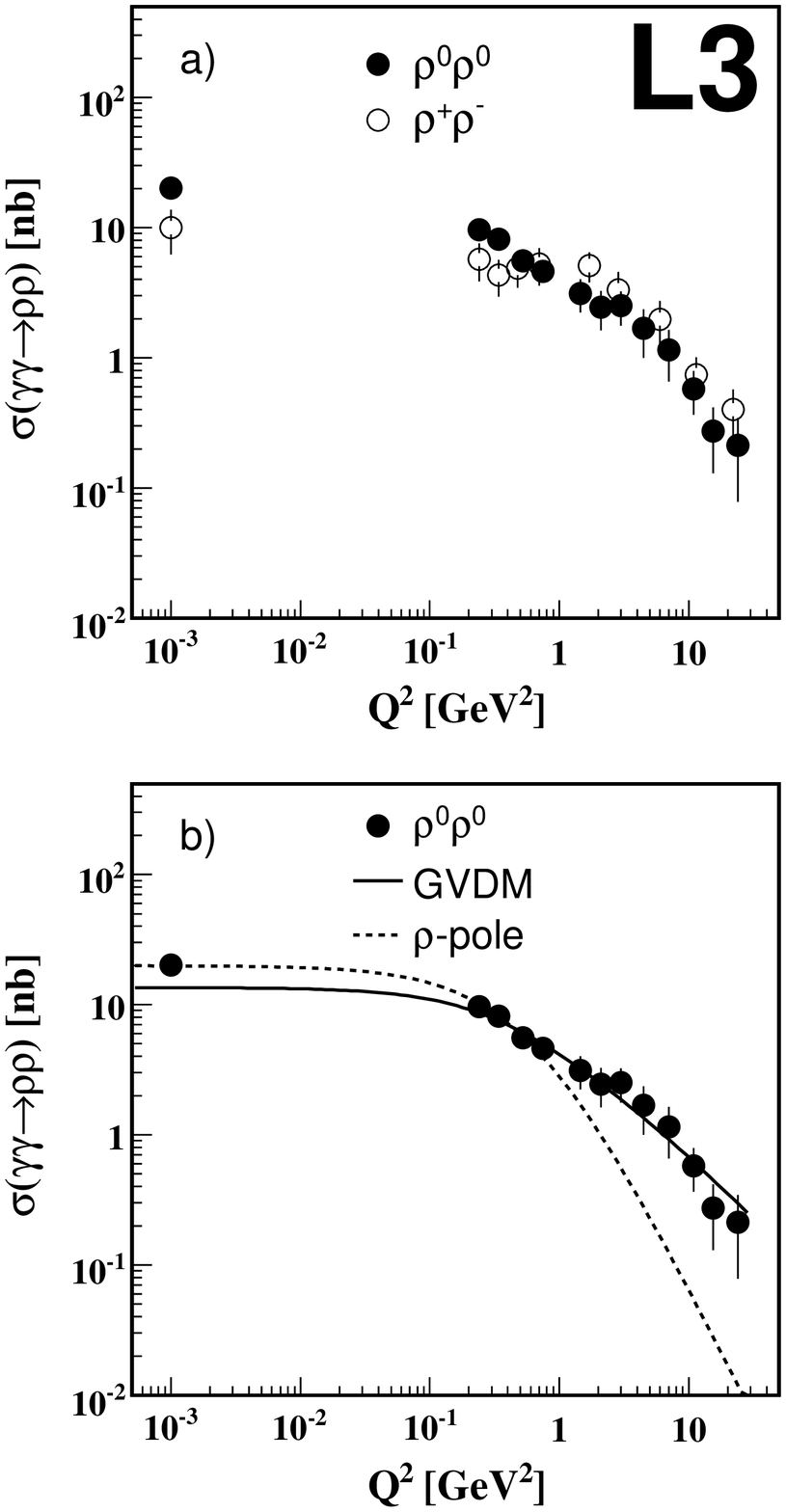}
\vspace{-0.3 cm }
\caption{a) The $\ggrzrz$ and $\ggrprm$ cross sections as a function of $Q^2$.
b) Comparison of the $\ggrzrz$ cross section as a function of
  $Q^2$ to a  GVDM and a simple  $\rho$-pole
  form-factor dependence, both fitted to previous L3 data 
  at higher $Q^2$~\cite{vsatch}.
The error bars show the statistical uncertainties.}
\label{fig:fig12}
\end{center}
\end{figure}

A spin-parity-helicity analysis of four-pion final states produced in quasi-real two-photon
collisions at LEP benefits from data 
 statistics an order of magnitude higher
than previous analyses. Several  characteristics of the
  $\ggrzrz$ and $\ggrprm$ processes, which were previously observed
\cite{ARGUS, ARGUS0}, are confirmed:
\begin{itemize}
  \item In both channels, the $(2^+,2)$ wave is dominant. Small but
significant $0^+$ and $0^-$ waves are also observed.
  \item The $\ggrzrz$ process has a high cross section extending from
  threshold to about 2 ~\GeV , while the cross section of the $\ggrprm$ process is low 
  in this range.
   In Figure~\ref{fig:fig11} the  mass spectra of the present results are compared to
  those we  obtained at higher $Q^2$~\cite{high00,highch,vsat00,vsatch}.
   The ratio 
  $$ R = \sum{\Delta \sigma_\mathrm{ee}(\rprm)} / \sum{ \Delta \sigma_\mathrm{ee}(\rzrz)} ,$$
where $\Delta \sigma_\mathrm{ee} =  \Delta{\cal L}_{\gamma\gamma}   \sigma_{\rm tot}(\gamma\gamma \ra \roro) $ and 
the sum is for  the region $ 1.1 \GeV \le  \wgg  \le 2.1 \GeV $, is found to be 
  $$ R = 0.42 \pm 0.05 \pm 0.09 \;\;\; \mathrm{for}\;\;  Q^2 \le 0.02 \GeV^2  .$$
 The first uncertainty is statistical and the second systematic, calculated assuming the systematic
  uncertainties for the two processes to be fully uncorrelated.
  This ratio increases with the photon virtuality. At higher  $Q^2$ we previously obtained:
  \begin{eqnarray}
  \begin{tabular}{cccccc}
  $R$ & = & $0.62  \pm 0.10  \pm 0.09 $ &for & $0.2\GeV^2   \le Q^2 \le   0.85 \GeV^2$& \cite{vsatch},  \nonumber \\  
  $R$ & = & $1.81 \pm 0.47 \pm 0.22 $ & for &  $1.2 \GeV^2  \le Q^2 \le  8.5 \phantom{0}\GeV^2$& \cite{vsatch},  \nonumber \\  
  $R$ & = & $   2.2\phantom{0} \pm 1.1\phantom{0} \pm 0.6\phantom{0} $& for &$  8.8 \GeV^2  \le Q^2 \le  30 \phantom{0}
  \;\GeV^2$& \cite{highch}.\nonumber 
  \end{tabular}
  \end{eqnarray}   
    These measurements  are consistent with the presence of an $s$-channel enhancement at low $\rzrz$ mass values
    which decreases rapidly with $Q^2$. If interpreted as an effect of $s$-channel resonances, the observed
    ratio between $\ggrprm$ and $\ggrzrz$ production implies the possible existence of an isospin-2 state~\cite{Li,Acha}.
    Such an interpretation of our data was recently presented  in Reference~\citen{Anikin}.
\item At higher masses, $ \wgg > 2 \GeV $, the  $\ggrprm$ and  $\ggrzrz$ cross sections are 
equal, within the experimental uncertainties. 
In both cases, the  cross section decreases rapidly for $ \wgg \sim 3 \GeV $. 
\end{itemize}
  The  $Q^2$-dependence of the two-photon cross section 
  is presented in Figure~\ref{fig:fig12} for the full mass-region  $ 1.1 \GeV \le  \wgg  \le 3. \GeV $.
  The $\rzrz$ cross section exceeds the $\rprm$ one at low  $Q^2$ while a cross-over is observed 
  in the vicinity of $Q^2 \simeq 1 \GeV ^2$. 
  A Generalised Vector Dominance fit, GVDM~\cite{gvdm}, which reproduces well all the mid-virtuality and high-virtuality
  data~\cite{vsatch}
  for the $\ggrzrz$ cross section,
  lies below the cross section value obtained at $\langle Q^2 \rangle = 0.001 \GeV^2$. 
  A $\rho $-pole fit, also presented in the Figure, better describes  the low-$Q^2$ region.

%
%
\bibliographystyle{l3stylem}

%
%
%
%
\newpage
\typeout{   }     
\typeout{Using author list for paper 287 -  }
\typeout{$Modified: Jul 15 2001 by smele $}
\typeout{!!!!  This should only be used with document option a4p!!!!}
\typeout{   }
%
%
%
%
%
%

\newcount\tutecount  \tutecount=0
\def\tutenum#1{\global\advance\tutecount by 1 \xdef#1{\the\tutecount}}
\def\tute#1{$^{#1}$}
\tutenum\aachen            
\tutenum\nikhef            
\tutenum\mich              
\tutenum\lapp              
\tutenum\basel             
\tutenum\lsu               
\tutenum\beijing           
\tutenum\bologna           
\tutenum\tata              
\tutenum\ne                
\tutenum\bucharest         
\tutenum\budapest          
\tutenum\mit               
\tutenum\panjab            
\tutenum\debrecen          
\tutenum\dublin            
\tutenum\florence          
\tutenum\cern              
\tutenum\wl                
\tutenum\geneva            
\tutenum\hamburg           
\tutenum\hefei             
\tutenum\lausanne          
\tutenum\lyon              
\tutenum\madrid            
\tutenum\florida           
\tutenum\milan             
\tutenum\moscow            
\tutenum\naples            
\tutenum\cyprus            
\tutenum\nymegen           
\tutenum\caltech           
\tutenum\perugia           
\tutenum\peters            
\tutenum\cmu               
\tutenum\potenza           
\tutenum\prince            
\tutenum\riverside         
\tutenum\rome              
\tutenum\salerno           
\tutenum\ucsd              
\tutenum\sofia             
\tutenum\korea             
\tutenum\taiwan            
\tutenum\tsinghua          
\tutenum\purdue            
\tutenum\psinst            
\tutenum\zeuthen           
\tutenum\eth               

{
\parskip=0pt
\noindent
{\bf The L3 Collaboration:}
\ifx\selectfont\undefined
 \baselineskip=10.8pt
 \baselineskip\baselinestretch\baselineskip
 \normalbaselineskip\baselineskip
 \ixpt
\else
 \fontsize{9}{10.8pt}\selectfont
\fi
\medskip
\tolerance=10000
\hbadness=5000
\raggedright
\hsize=162truemm\hoffset=0mm
\def\r{\rlap,}
\noindent

P.Achard\r\tute\geneva\ 
O.Adriani\r\tute{\florence}\ 
M.Aguilar-Benitez\r\tute\madrid\ 
J.Alcaraz\r\tute{\madrid}\ 
G.Alemanni\r\tute\lausanne\
J.Allaby\r\tute\cern\
A.Aloisio\r\tute\naples\ 
M.G.Alviggi\r\tute\naples\
H.Anderhub\r\tute\eth\ 
V.P.Andreev\r\tute{\lsu,\peters}\
% ...
\newpage
\begin{list}{A}{\itemsep=0pt plus 0pt minus 0pt\parsep=0pt plus 0pt minus 0pt
                \topsep=0pt plus 0pt minus 0pt}
\item[\aachen]
 III. Physikalisches Institut, RWTH, D-52056 Aachen, Germany$^{\S}$
\item[\nikhef] National Institute for High Energy Physics, NIKHEF, 
     and University of Amsterdam, NL-1009 DB Amsterdam, The Netherlands
\item[\mich] University of Michigan, Ann Arbor, MI 48109, USA
\item[\lapp] Laboratoire d'Annecy-le-Vieux de Physique des Particules, 
     LAPP,IN2P3-CNRS, BP 110, F-74941 Annecy-le-Vieux CEDEX, France
\item[\basel] Institute of Physics, University of Basel, CH-4056 Basel,
     Switzerland
\item[\lsu] Louisiana State University, Baton Rouge, LA 70803, USA
\item[\beijing] Institute of High Energy Physics, IHEP, 
  100039 Beijing, China$^{\triangle}$ 
\item[\bologna] University of Bologna and INFN-Sezione di Bologna, 
     I-40126 Bologna, Italy
\item[\tata] Tata Institute of Fundamental Research, Mumbai (Bombay) 400 005, India
\item[\ne] Northeastern University, Boston, MA 02115, USA
\item[\bucharest] Institute of Atomic Physics and University of Bucharest,
     R-76900 Bucharest, Romania
\item[\budapest] Central Research Institute for Physics of the 
     Hungarian Academy of Sciences, H-1525 Budapest 114, Hungary$^{\ddag}$
\item[\mit] Massachusetts Institute of Technology, Cambridge, MA 02139, USA
\item[\panjab] Panjab University, Chandigarh 160 014, India
\item[\debrecen] KLTE-ATOMKI, H-4010 Debrecen, Hungary$^\P$
\item[\dublin] UCD School of Physics, University College Dublin, 
 Belfield, Dublin 4, Ireland
\item[\florence] INFN Sezione di Firenze and University of Florence, 
     I-50125 Florence, Italy
\item[\cern] European Laboratory for Particle Physics, CERN, 
     CH-1211 Geneva 23, Switzerland
\item[\wl] World Laboratory, FBLJA  Project, CH-1211 Geneva 23, Switzerland
\item[\geneva] University of Geneva, CH-1211 Geneva 4, Switzerland
\item[\hamburg] University of Hamburg, D-22761 Hamburg, Germany
\item[\hefei] Chinese University of Science and Technology, USTC,
      Hefei, Anhui 230 029, China$^{\triangle}$
\item[\lausanne] University of Lausanne, CH-1015 Lausanne, Switzerland
\item[\lyon] Institut de Physique Nucl\'eaire de Lyon, 
     IN2P3-CNRS,Universit\'e Claude Bernard, 
     F-69622 Villeurbanne, France
\item[\madrid] Centro de Investigaciones Energ{\'e}ticas, 
     Medioambientales y Tecnol\'ogicas, CIEMAT, E-28040 Madrid,
     Spain${\flat}$ 
\item[\florida] Florida Institute of Technology, Melbourne, FL 32901, USA
\item[\milan] INFN-Sezione di Milano, I-20133 Milan, Italy
\item[\moscow] Institute of Theoretical and Experimental Physics, ITEP, 
     Moscow, Russia
\item[\naples] INFN-Sezione di Napoli and University of Naples, 
     I-80125 Naples, Italy
\item[\cyprus] Department of Physics, University of Cyprus,
     Nicosia, Cyprus
\item[\nymegen] Radboud University and NIKHEF, 
     NL-6525 ED Nijmegen, The Netherlands
\item[\caltech] California Institute of Technology, Pasadena, CA 91125, USA
\item[\perugia] INFN-Sezione di Perugia and Universit\`a Degli 
     Studi di Perugia, I-06100 Perugia, Italy   
\item[\peters] Nuclear Physics Institute, St. Petersburg, Russia
\item[\cmu] Carnegie Mellon University, Pittsburgh, PA 15213, USA
\item[\potenza] INFN-Sezione di Napoli and University of Potenza, 
     I-85100 Potenza, Italy
\item[\prince] Princeton University, Princeton, NJ 08544, USA
\item[\riverside] University of Californa, Riverside, CA 92521, USA
\item[\rome] INFN-Sezione di Roma and University of Rome, ``La Sapienza",
     I-00185 Rome, Italy
\item[\salerno] University and INFN, Salerno, I-84100 Salerno, Italy
\item[\ucsd] University of California, San Diego, CA 92093, USA
\item[\sofia] Bulgarian Academy of Sciences, Central Lab.~of 
     Mechatronics and Instrumentation, BU-1113 Sofia, Bulgaria
\item[\korea]  The Center for High Energy Physics, 
     Kyungpook National University, 702-701 Taegu, Republic of Korea
\item[\taiwan] National Central University, Chung-Li, Taiwan, China
\item[\tsinghua] Department of Physics, National Tsing Hua University,
      Taiwan, China
\item[\purdue] Purdue University, West Lafayette, IN 47907, USA
\item[\psinst] Paul Scherrer Institut, PSI, CH-5232 Villigen, Switzerland
\item[\zeuthen] DESY, D-15738 Zeuthen, Germany
\item[\eth] Eidgen\"ossische Technische Hochschule, ETH Z\"urich,
     CH-8093 Z\"urich, Switzerland
\item[\S]  Supported by the German Bundesministerium 
        f\"ur Bildung, Wissenschaft, Forschung und Technologie.
\item[\ddag] Supported by the Hungarian OTKA fund under contract
numbers T019181, F023259 and T037350.
\item[\P] Also supported by the Hungarian OTKA fund under contract
  number T026178.
\item[$\flat$] Supported also by the Comisi\'on Interministerial de Ciencia y 
        Tecnolog{\'\i}a.
\item[$\sharp$] Also supported by CONICET and Universidad Nacional de La Plata,
        CC 67, 1900 La Plata, Argentina.
\item[$\triangle$] Supported by the National Natural Science
  Foundation of China.
\item[$\dagger$] Deceased.
\end{list}
}
\vfill


\newpage
\begin{sidewaystable}
  \begin{center}

    \hbox{\vbox{
\tabskip0pt
\offinterlineskip
\hskip-1cm\halign {\vrule#&&\strut\ \hfil$#$\hfil\ &\vrule#\cr
\noalign{\hrule}
height3pt&\omit&&\omit&&\omit&&\omit&&\omit&&\omit&&\omit&&\omit&&\omit&\omit\vrule\hskip.6pt\vrule&\omit&\cr
&W_{\gamma\gamma}\; [\GeV]&&N&& \int \d\mathcal{L}_{\gaga}  [10^{-3}]&&\varepsilon_\mathrm{trg}\; [\%]&&\varepsilon \; [\%]&&4\pi\;  [\mathrm{nb}]&&0^+
\;[\mathrm{nb}]&&0^- \;[\mathrm{nb}]&&(2^+,2) \;[\mathrm{nb}]&\omit\vrule\hskip.6pt\vrule&\sigma_{\rm tot}(\gamma\gamma\to\rho ^0\rho ^0)\; [\mathrm{nb}]&\cr
height3pt&\omit&&\omit&&\omit&&\omit&&\omit&&\omit&&\omit&&\omit&&\omit&\omit\vrule\hskip.6pt\vrule&\omit&\cr
\noalign{\hrule}height3pt&\omit&&\omit&&\omit&&\omit&&\omit&&\omit&&\omit&&\omit&&\omit&\omit\vrule\hskip.6pt\vrule&\omit&\cr
&\omit\hfill1.00--1.10\hfill&&\phantom{0}376&&4.06&&94.2&&1.8&&\phantom{0}
3.8\pm0.7\pm0.1&&\phantom{0} 0.6\pm0.4\pm0.1&& -- &&\phantom{0} 1.4\pm0.5\pm0.1&\omit\vrule\hskip.6pt\vrule&\phantom{0} 2.1\pm0.7\pm0.1&\cr
&\omit\hfill1.10--1.20\hfill&& 1099&&3.58&&94.2&&2.7&&\phantom{0}
3.7\pm0.7\pm0.1&&\phantom{0} 0.7\pm0.4\pm0.2&& -- &&\phantom{0} 6.2\pm0.6\pm0.2&\omit\vrule\hskip.6pt\vrule&\phantom{0} 6.9\pm0.7\pm0.2&\cr
&\omit\hfill1.20--1.30\hfill&& 4513&&3.20&&95.3&&3.5&&\phantom{0} 5.3\pm1.0\pm0.4&&\phantom{0} 5.1\pm1.1\pm0.4&&\phantom{0} 0.8\pm0.6\pm0.1&&23.2\pm1.4\pm1.8&\omit\vrule\hskip.6pt\vrule&29.1\pm1.9\pm2.2&\cr
&\omit\hfill1.30--1.40\hfill&& 7717&&2.87&&95.3&&4.2&&16.3\pm1.3\pm1.0&&\phantom{0} 7.7\pm1.3\pm0.5&&\phantom{0} 1.4\pm0.7\pm0.1&&30.5\pm1.6\pm1.9&\omit\vrule\hskip.6pt\vrule&39.6\pm2.2\pm2.5&\cr
&\omit\hfill1.40--1.50\hfill&& 9084&&2.60&&95.3&&4.8&&18.2\pm1.2\pm1.0&&13.7\pm1.5\pm0.8&&\phantom{0} 1.8\pm0.7\pm0.1&&31.7\pm1.7\pm1.8&\omit\vrule\hskip.6pt\vrule&47.1\pm2.4\pm2.7&\cr
&\omit\hfill1.50--1.60\hfill&& 8397&&2.37&&95.8&&5.4&&19.8\pm1.2\pm2.2&&\phantom{0} 9.8\pm1.5\pm1.1&&\phantom{0} 4.3\pm0.9\pm0.5&&34.5\pm2.0\pm3.8&\omit\vrule\hskip.6pt\vrule&48.6\pm2.6\pm5.3&\cr
&\omit\hfill1.60--1.70\hfill&& 7910&&2.17&&95.8&&5.9&&19.3\pm1.2\pm2.1&&\phantom{0} 5.9\pm1.3\pm0.7&&\phantom{0} 2.1\pm0.7\pm0.2&&35.6\pm1.8\pm3.9&\omit\vrule\hskip.6pt\vrule&43.7\pm2.4\pm4.8&\cr
&\omit\hfill1.70--1.80\hfill&& 6671&&2.00&&96.2&&6.3&&19.0\pm1.2\pm1.0&&\phantom{0} 7.7\pm1.3\pm0.4&&\phantom{0} 4.8\pm0.8\pm0.2&&26.4\pm1.6\pm1.3&\omit\vrule\hskip.6pt\vrule&39.0\pm2.2\pm2.0&\cr
&\omit\hfill1.80--1.90\hfill&& 5643&&1.85&&96.2&&6.7&&20.7\pm1.3\pm1.7&&\phantom{0} 4.9\pm1.1\pm0.4&&\phantom{0} 7.4\pm0.9\pm0.6&&23.6\pm1.5\pm1.9&\omit\vrule\hskip.6pt\vrule&35.9\pm2.1\pm2.9&\cr
&\omit\hfill1.90--2.00\hfill&& 4965&&1.72&&96.2&&7.1&&27.8\pm1.6\pm3.3&&\phantom{0} 7.3\pm1.2\pm0.9&&\phantom{0} 4.8\pm0.8\pm0.6&&15.0\pm1.3\pm1.8&\omit\vrule\hskip.6pt\vrule&27.1\pm1.9\pm3.2&\cr
&\omit\hfill2.00--2.10\hfill&& 4004&&1.60&&96.4&&7.4&&26.2\pm1.6\pm6.0&&\phantom{0} 9.5\pm1.3\pm2.2&&\phantom{0} 4.0\pm0.7\pm0.9&&\phantom{0} 7.9\pm1.1\pm1.8&\omit\vrule\hskip.6pt\vrule&21.3\pm1.8\pm4.9&\cr
&\omit\hfill2.10--2.20\hfill&& 3118&&1.49&&96.4&&7.7&&24.4\pm1.5\pm8.9&&\phantom{0} 4.8\pm1.0\pm1.8&&\phantom{0} 2.0\pm0.5\pm0.7&&\phantom{0} 7.2\pm1.0\pm2.6&\omit\vrule\hskip.6pt\vrule&13.9\pm1.5\pm5.1&\cr
&\omit\hfill2.20--2.30\hfill&& 2366&&1.40&&96.2&&7.9&&21.0\pm1.4\pm8.5&&\phantom{0} 2.2\pm0.7\pm0.9&&\phantom{0} 1.8\pm0.5\pm0.7&&\phantom{0} 5.2\pm0.9\pm2.1&\omit\vrule\hskip.6pt\vrule&\phantom{0} 9.2\pm1.3\pm3.7&\cr
&\omit\hfill2.30--2.40\hfill&& 1763&&1.31&&96.2&&8.1&&17.3\pm1.3\pm9.7&&\phantom{0} 1.6\pm0.6\pm0.9&&\phantom{0} 2.6\pm0.6\pm1.5&&\phantom{0} 2.6\pm0.7\pm1.4&\omit\vrule\hskip.6pt\vrule&\phantom{0} 6.8\pm1.1\pm3.8&\cr
&\omit\hfill2.40--2.50\hfill&& 1450&&1.24&&96.2&&8.4&&15.0\pm1.1\pm8.9&&\phantom{0} 2.0\pm0.6\pm1.2&&\phantom{0} 1.7\pm0.5\pm1.0&&\phantom{0} 1.4\pm0.7\pm0.8&\omit\vrule\hskip.6pt\vrule&\phantom{0} 5.1\pm1.0\pm3.0&\cr
&\omit\hfill2.50--2.60\hfill&& 1137&&1.17&&95.8&&8.6&&12.1\pm1.1\pm7.3&&\phantom{0} 2.0\pm0.7\pm1.2&&\phantom{0} 1.0\pm0.4\pm0.6&&\phantom{0} 2.1\pm0.8\pm1.2&\omit\vrule\hskip.6pt\vrule&\phantom{0} 5.1\pm1.1\pm3.1&\cr
&\omit\hfill2.60--2.70\hfill&&\phantom{0}878&&1.10&&95.8&&8.8&&10.7\pm1.0\pm6.8&&\phantom{0} 1.4\pm0.5\pm0.9&&\phantom{0} 1.5\pm0.4\pm1.0&&--&\omit\vrule\hskip.6pt\vrule&\phantom{0} 2.9\pm0.6\pm1.9&\cr
&\omit\hfill2.70--2.80\hfill&&\phantom{0}672&&1.05&&96.5&&8.9&&\phantom{0} 8.5\pm0.9\pm2.3&&\phantom{0} 1.1\pm0.3\pm0.3&&-- &&\phantom{0} 1.1\pm0.3\pm0.3&\omit\vrule\hskip.6pt\vrule&\phantom{0} 2.2\pm0.5\pm0.6&\cr
&\omit\hfill2.80--2.90\hfill&&\phantom{0}545&&0.99&&96.5&&9.1&&\phantom{0}
7.3\pm0.8\pm1.1&&\phantom{0} 0.7\pm0.2\pm0.1&& -- &&\phantom{0} 1.4\pm0.3\pm0.2&\omit\vrule\hskip.6pt\vrule&\phantom{0} 2.1\pm0.4\pm0.3&\cr
&\omit\hfill2.90--3.00\hfill&&\phantom{0}467&&0.94&&96.5&&9.3&&\phantom{0}
6.4\pm0.8\pm0.1&&\phantom{0} 1.5\pm0.5\pm0.1&& -- && -- &\omit\vrule\hskip.6pt\vrule&\phantom{0} 1.7\pm0.6\pm0.1&\cr
height3pt&\omit&&\omit&&\omit&&\omit&&\omit&&\omit&&\omit&&\omit&&\omit&\omit\vrule\hskip.6pt\vrule&\omit&\cr
\noalign{\hrule}
height3pt&\omit&&\omit&&\omit&&\omit&&\omit&&\omit&&\omit&&\omit&&\omit&\omit\vrule\hskip.6pt\vrule&\omit&\cr
&\omit\hfill1.00--3.00\hfill&&72775&&38.72&&-&&-&&14.2\pm1.1\pm2.7&& 4.9\pm0.9\pm0.6&& 2.0\pm0.5\pm0.3&&15.4\pm1.1\pm1.5&\omit\vrule\hskip.6pt\vrule&22.3\pm1.6\pm2.5&\cr
height3pt&\omit&&\omit&&\omit&&\omit&&\omit&&\omit&&\omit&&\omit&&\omit&\omit\vrule\hskip.6pt\vrule&\omit&\cr
\noalign{\hrule}}    
    }}
    \caption{Cross section measurements and fit results for $\ggpppp$ for different $\wgg$ intervals.
  $N$ is the number of events in a bin, $\int \d\mathcal{L}_{\gaga}$ the
two-photon luminosity function, $\varepsilon_\mathrm{ trg}$ 
the  trigger efficiency and $\varepsilon$ the selection
efficiency. The cross sections for the background, $ 4 \pi$, and for  the
different spin-helicity waves are given, along with the total $\gaga\to\rz\rz$ cross
section. A double dash indicates that no significant contribution to
  the fit is observed. The first uncertainties are statistical, the second
systematic.} 
    \label{tab:xsec0}  
  \end{center}
\end{sidewaystable}

\begin{sidewaystable}
   \begin{center}
      \hbox{\hfill\vbox{
\tabskip0pt
\offinterlineskip
\hskip-1cm\halign {\vrule#&&\strut\ \hfil$#$\hfil\ &\vrule#\cr
\noalign{\hrule}
height3pt&\omit&&\omit&&\omit&&\omit&&\omit&&\omit&&\omit&&\omit&&\omit&\omit\vrule\hskip.6pt\vrule&\omit&\cr
&W_{\gamma\gamma} \; [\GeV]&&N&& \int \d{\mathcal{L}}_{\gaga} [10^{-3}]&&\varepsilon_\mathrm{trg}\; [\%]&&\varepsilon \;
[\%]&&4\pi \;[\mathrm{nb}]&&0^+ \;[\mathrm{nb}]&&0^-\; [\mathrm{nb}]&&(2^+,2) \;
[\mathrm{nb}]&\omit\vrule\hskip.6pt\vrule&\sigma_{\rm tot}(\gamma\gamma\to\rho^+\rho^-)\; [\mathrm{nb}]&\cr
height3pt&\omit&&\omit&&\omit&&\omit&&\omit&&\omit&&\omit&&\omit&&\omit&\omit\vrule\hskip.6pt\vrule&\omit&\cr
\noalign{\hrule}height3pt&\omit&&\omit&&\omit&&\omit&&\omit&&\omit&&\omit&&\omit&&\omit&\omit\vrule\hskip.6pt\vrule&\omit&\cr
&\omit\hfill1.00--1.20\hfill&&\phantom{0}111&&7.64&&66.4&&0.3&&\phantom{0} 7.6\pm2.2\pm1.1&&\phantom{0} 0.6\pm1.1\pm0.1&&\phantom{0} 0.6\pm0.8\pm0.1&&\phantom{0} 1.0\pm1.2\pm0.1&\omit\vrule\hskip.6pt\vrule&\phantom{0} 2.2\pm1.8\pm0.3&\cr
&\omit\hfill1.20--1.40\hfill&&\phantom{0}526&&6.07&&63.5&&0.5&&20.1\pm3.3\pm2.8&&\phantom{0} 3.7\pm2.7\pm0.5&&\phantom{0} 1.8\pm1.3\pm0.2&&\phantom{0} 7.9\pm2.7\pm1.1&\omit\vrule\hskip.6pt\vrule&13.4\pm4.0\pm1.8&\cr
&\omit\hfill1.40--1.60\hfill&&\phantom{0}839&&4.97&&65.0&&0.8&&30.7\pm3.4\pm5.5&&\phantom{0} 1.5\pm2.1\pm0.3&&\phantom{0} 4.5\pm1.3\pm0.8&&\phantom{0} 5.0\pm2.0\pm0.9&\omit\vrule\hskip.6pt\vrule&10.9\pm3.2\pm2.0&\cr
&\omit\hfill1.60--1.80\hfill&& 1160&&4.17&&65.0&&1.0&&30.8\pm3.5\pm5.5&&\phantom{0} 4.8\pm2.2\pm0.9&&\phantom{0} 1.4\pm1.0\pm0.3&&12.3\pm2.3\pm2.2&\omit\vrule\hskip.6pt\vrule&18.6\pm3.3\pm3.3&\cr
&\omit\hfill1.80--2.00\hfill&& 1205&&3.56&&59.9&&1.3&&32.2\pm3.8\pm8.6&&\phantom{0} 3.8\pm2.3\pm1.0&&\phantom{0} 2.9\pm1.5\pm0.8&&19.2\pm3.0\pm5.1&\omit\vrule\hskip.6pt\vrule&25.9\pm4.1\pm6.9&\cr
&\omit\hfill2.00--2.20\hfill&&
1161&&3.09&&63.4&&1.6&&34.0\pm3.7\pm8.3&&\phantom{0} 8.5\pm2.2\pm2.1&&
-- &&\phantom{0} 9.1\pm2.1\pm2.2&\omit\vrule\hskip.6pt\vrule&17.7\pm3.2\pm4.3&\cr
&\omit\hfill2.20--2.40\hfill&&\phantom{0}823&&2.71&&64.4&&1.8&&27.9\pm3.3\pm12&&\phantom{0} 2.6\pm1.4\pm1.1&&\phantom{0} 2.6\pm1.2\pm1.1&&\phantom{0}
3.5\pm1.6\pm1.5&\omit\vrule\hskip.6pt\vrule&\phantom{0} 8.6\pm2.5\pm3.7&\cr
&\omit\hfill2.40--2.60\hfill&&\phantom{0}540&&2.41&&62.8&&2.1&&17.2\pm2.5\pm7.4&&\phantom{0} 1.7\pm1.1\pm0.7&&\phantom{0} 1.0\pm0.7\pm0.4&&\phantom{0} 2.7\pm1.3\pm1.1&\omit\vrule\hskip.6pt\vrule&\phantom{0} 5.4\pm1.8\pm2.3&\cr
&\omit\hfill2.60--2.80\hfill&&\phantom{0}336&&2.15&&62.8&&2.3&&12.0\pm2.0\pm3.8&&\phantom{0}
1.6\pm0.9\pm0.5&&\phantom{0} 2.4\pm1.1\pm0.8&& --&\omit\vrule\hskip.6pt\vrule&\phantom{0} 4.3\pm1.8\pm1.4&\cr
&\omit\hfill2.80--3.00\hfill&&\phantom{0}231&&1.94&&68.7&&2.6&&\phantom{0} 7.2\pm1.4\pm1.4&&\phantom{0} 1.3\pm0.6\pm0.3&&--&&\phantom{0} 1.4\pm0.6\pm0.3&\omit\vrule\hskip.6pt\vrule&\phantom{0} 2.7\pm0.8\pm0.5&\cr
height3pt&\omit&&\omit&&\omit&&\omit&&\omit&&\omit&&\omit&&\omit&&\omit&\omit\vrule\hskip.6pt\vrule&\omit&\cr
\noalign{\hrule}
height3pt&\omit&&\omit&&\omit&&\omit&&\omit&&\omit&&\omit&&\omit&&\omit&\omit\vrule\hskip.6pt\vrule&\omit&\cr
&\omit\hfill1.00--3.00\hfill&& 6932&&38.72&&-&&-&&21.7\pm3.0\pm5.8&& 2.9\pm1.8\pm0.7&& 1.8\pm1.0\pm0.5&& 6.4\pm1.9\pm1.5&\omit\vrule\hskip.6pt\vrule&11.0\pm2.8\pm2.7&\cr
height3pt&\omit&&\omit&&\omit&&\omit&&\omit&&\omit&&\omit&&\omit&&\omit&\omit\vrule\hskip.6pt\vrule&\omit&\cr
\noalign{\hrule}}       
     }}
      \caption{Cross section measurement and fit results for $\ggppzppz$ for different $\wgg$ intervals.
  $N$ is the number of events in a bin, $\int \d\mathcal{L}_{\gaga} $ the
two-photon luminosity function, $\varepsilon_\mathrm{ trg}$ 
the  trigger efficiency and $\varepsilon$ the selection
efficiency. The cross sections for the background, $ 4 \pi$, and  for the
different spin-helicity waves are given together with the total  $\ggrprm$ cross
section.  A double dash indicates that no significant contribution to
  the fit is observed. The first uncertainties are statistical, the second
systematic.
} 
      \label{tab:xsec1}
   \end{center}
\end{sidewaystable}

\newpage

\newpage 

\newpage
\vspace{-0.3 cm }

\newpage

\vspace{-2 cm }

\newpage

\vskip-2cm

\begin{thebibliography}{99}
%
%
\def\PL{Phys. Lett. }
\def\PRt{Phys. Rep. }
\def\ZP{Z. Phys. }
\def\PR{Phys. Rev. }
\def\PRL{Phys. Rev. Lett. }
\def\NIM{Nucl. Inst. Meth. }
\bibitem{TASSO} TASSO Collab., R.~Brandelik \etal, \PL {\bf  B 97}
  (1980) 448;\\
   MARK II Collab., D.L.~Burke \etal, \PL {\bf B 103} (1981)
  153;  \\
  CELLO Collab., H.-J.~Behrend \etal, \ZP {\bf C 21} (1984) 205;\\
 PLUTO Collab., Ch.~Berger \etal, \ZP {\bf C 38} (1988) 521; \\
TPC/Two-Gamma Collab., H.~Aihara \etal, \PR {\bf D 37} (1988) 28.

\bibitem{TASSO1} TASSO Collab., M.~Althoff \etal, \ZP {\bf C 16} (1982) 13.
\bibitem{ARGUS} ARGUS Collab., H.~Albrecht \etal, \ZP {\bf C 50} (1991) 1.
%
%
\bibitem{ARGUS1} ARGUS Collab., H.~Albrecht \etal, \PL {\bf B 217} (1989) 205. 
\bibitem{ARGUS0} ARGUS Collab., H.~Albrecht \etal, \PL {\bf B 267} (1991) 535.
%
%
\bibitem{Rosner} J.L. Rosner, \PR {\bf D 70} (2004) 034028 and references therein.
\bibitem{Alex} G.~Alexander, U.~Maor, P.G.~Williams, \PR {\bf D 26} (1982) 1198;\\
G.~Alexander, A.~Levy, U.~Maor,  \ZP {\bf C 30} (1986) 65.
\bibitem{Li} B.A. Li and  K.F.~Liu, \PL {\bf B 118} (1982) 435;\\
B.A. Li and  K.F.~Liu, \PRL {\bf 51} (1983) 1510;\\
B.A. Li and  K.F.~Liu,  \PR {\bf D 30} (1984) 613;\\
B.A. Li and  K.F.~Liu, \PRL {\bf 58} (1987) 2288.
\bibitem{Acha} 
 N.N.~Achasov \etal, \PL {\bf  B 108} (1982) 134;\\
 N.N.~Achasov \etal, \ZP {\bf C 16} (1982) 55;\\
 N.N.~Achasov \etal, \ZP {\bf C 27} (1985) 99;\\
 N.N.~Achasov \etal, \PL {\bf B 203} (1988) 309.
\bibitem{L3D} L3 Collab., B.~Adeva \etal, \NIM {\bf A 289} (1990) 35;\\
L3 Collab., O.~Adriani \etal, \PRt {\bf 236} (1993) 1;\\
M.~Acciarri \etal, \NIM {\bf A 351} (1994) 30;\\
M.~Chemarin \etal, \NIM {\bf A 349} (1994) 345;\\
I.C.~Brock \etal, \NIM {\bf A 381} (1996) 236;\\
A.~Adam \etal, \NIM {\bf A 383} (1996) 342.
 \bibitem{high00}
 L3 Collab., P.~Achard \etal, \PL {\bf B 568} (2003) 11.
\bibitem{highch}
 L3 Collab., P.~Achard \etal, \PL {\bf B 597} (2004) 26.
\bibitem{vsat00}
 L3 Collab., P.~Achard \etal, \PL {\bf B 604} (2004) 48. 
\bibitem{vsatch}
 L3 Collab., P.~Achard \etal, \PL {\bf B 615} (2005) 19.
\bibitem{EGPC} F. L. Linde ``Charm Production in Two-Photon Collisions'', 
Ph. D. Thesis, Rijksuniversiteit Leiden, (1988).
\bibitem{Bud} V.M. Budnev \etal, \PRt  {\bf 15} (1974) 181.
\bibitem{GEANT} R. Brun \etal, preprint CERN DD/EE/84-1 (1984), revised 1987.
\bibitem{GEISHA} H. Fesefeldt, RWTH Aachen report PITHA 85/2 (1985).

\bibitem{L3T} P.~B\'en\'e et al., \NIM {\bf A 306} (1991) 150.
\bibitem{L3IT} D.~Haas et al., \NIM {\bf A 420} (1999) 101.
\bibitem{Jack} J.D.~Jackson, Nuovo Cimento {\bf 34} (1964) 1644.    
\bibitem{Landau} L.D.~Landau, Dokl. Akad. Nauk (USSR) {\bf 60} (1948) 207; English summary
in Phys. Abstracts {\bf A 52} (1949) 125;\\
C. M. Yang, \PR {\bf 77} (1950) 242.
\bibitem{Anikin}
I. V. Anikin, B. Pire and O. V. Teryaev,  \PL{\bf B 626} (2005) 86.
 \bibitem{gvdm} J.J. Sakurai and D. Schildknecht,  \PL{\bf B40} (1972)  121;\\
I.F. Ginzburg and V.G. Serbo,  \PL{\bf B 109} (1982) 231.


\end{thebibliography}
\end{document}